%% file: sample-manuscript.tex
\PassOptionsToPackage{table}{xcolor}
\documentclass[acmsmall]{acmart}   %

\usepackage{amsmath}
\usepackage{graphicx}
\usepackage{subcaption}
\usepackage{multirow}
\usepackage{array}
\newcolumntype{U}{>{\centering\arraybackslash}p{0.02\textwidth}}
\usepackage{stfloats}
\usepackage{booktabs}
\usepackage{enumitem}
\usepackage{xspace}
\usepackage{pifont}
\usepackage{tabularx}
\usepackage{hyphenat}
\usepackage{wrapfig}
\usepackage{makecell}
\usepackage{listings}
\usepackage{fancyvrb}
\usepackage[most]{tcolorbox}
\usepackage{etoc}
\newcommand{\framework}{\textbf{\texttt{TSN4PI}}\xspace} %
\newcommand{\pidn}{\textbf{\texttt{PIDN}}\xspace} %
\newcommand{\pipn}{\textbf{\texttt{PIPN}}\xspace} %
\definecolor{lightgray}{gray}{0.90}
\definecolor{lightpink}{rgb}{1.0, 0.9, 0.9}
\definecolor{lightblue}{rgb}{0.9, 0.95, 1.0} %
\definecolor{codegreen}{rgb}{0,0.6,0}
\definecolor{codegray}{rgb}{0.5,0.5,0.5}
\definecolor{codepurple}{rgb}{0.58,0,0.82}
\definecolor{backcolour}{rgb}{0.95,0.95,0.92}

\lstdefinelanguage{json}{
  basicstyle=\ttfamily\footnotesize,
  string=[s]{"}{"},
  commentstyle=\color{codegreen},
  keywordstyle=\color{blue},
  stringstyle=\color{codepurple},
  numbers=none,
  showstringspaces=false,
  breaklines=true,
  frame=l,
  morekeywords={true,false,null},
  literate=
    *{0}{{{\color{codegray}0}}}1
    {1}{{{\color{codegray}1}}}1
    {2}{{{\color{codegray}2}}}1
    {3}{{{\color{codegray}3}}}1
    {4}{{{\color{codegray}4}}}1
    {5}{{{\color{codegray}5}}}1
    {6}{{{\color{codegray}6}}}1
    {7}{{{\color{codegray}7}}}1
    {8}{{{\color{codegray}8}}}1
    {9}{{{\color{codegray}9}}}1
}

\lstdefinestyle{jsonstyle}{
  language=json,
  backgroundcolor=\color{backcolour},
  commentstyle=\color{codegreen},
  keywordstyle=\color{blue},
  numberstyle=\tiny\color{codegray},
  stringstyle=\color{codepurple},
  basicstyle=\ttfamily\footnotesize,
  breakatwhitespace=false,
  breaklines=true,
  captionpos=b,
  keepspaces=true,
  showspaces=false,
  showstringspaces=false,
  showtabs=false,
  tabsize=2,
  frame=l,
  xleftmargin=2mm,
}
\setlist[itemize]{
  topsep=2pt,     %
  itemsep=1.5pt,  %
  parsep=0pt,     %
  partopsep=1pt,  %
  left=1.2em      %
}

\newcommand{\smref}[1]{#1}
\newcommand{\cmark}{\ding{51}}
\newcommand{\xmark}{\ding{55}}
\newcommand{\pmstd}[2]{#1{\scriptsize$\,\pm\,$#2}}
\newcommand{\best}[1]{\textbf{#1}}
\newcommand{\secbest}[1]{\underline{#1}}

\setcopyright{acmlicensed}
\copyrightyear{2026}
\acmYear{2026}
\acmDOI{XXXXXXX.XXXXXXX}

\acmJournal{TIST}

\AtBeginDocument{%
  \setlength{\abovedisplayskip}{5pt plus 2pt minus 2pt}%
  \setlength{\belowdisplayskip}{5pt plus 2pt minus 2pt}%
  \setlength{\abovedisplayshortskip}{3pt plus 2pt}%
  \setlength{\belowdisplayshortskip}{4pt plus 2pt minus 2pt}%
}

\begin{document}

\title{Against Political Polarization: A Unified Framework for Tracing Evolving Political Ideologies on Social Media}

\author{Yijie Xu}
\email{yxu409@connect.hkust-gz.edu.cn}
\orcid{0009-0008-4529-2701}
\affiliation{%
  \institution{Thrust of Artificial Intelligence, The Hong Kong University of Science and Technology (Guangzhou)}
  \city{Guangzhou}
  \country{China}
}

\author{Chao Wang}
\authornote{Corresponding authors.}
\email{wangchaoai@ustc.edu.cn}
\affiliation{%
  \institution{School of Artificial Intelligence and Data Science, University of Science and Technology of China}
  \city{Hefei}
  \country{China}
}

\author{Hui Xiong}
\authornotemark[1]
\email{xionghui@ust.hk}
\affiliation{%
  \institution{Thrust of Artificial Intelligence, The Hong Kong University of Science and Technology (Guangzhou)}
  \city{Guangzhou}
  \country{China}
}
\affiliation{%
  \institution{Department of Computer Science and Engineering, The Hong Kong University of Science and Technology}
  \city{Hong Kong}
  \country{Hong Kong SAR, China}
}
\renewcommand{\shortauthors}{Xu, et al.}

\begin{abstract}
The rapid growth of social media has greatly influenced political discourse, highlighting the need to understand individual political ideologies and their temporal dynamics. This task faces challenges such as data scarcity, abundant non-political content, costly and bias-prone manual annotation, and difficulty in modeling future ideological inclinations.
To address these issues, we propose \framework, a unified framework for tracking the evolution of political ideologies on social media. It includes two core modules. The \textbf{\pidn} uses large language models with style transfer and unsupervised domain adaptation to enable robust ideology detection and filter irrelevant content from noisy, cross-domain data. The \textbf{\pipn} employs temporal graph neural networks to predict future ideological shifts, enabling comprehensive analysis of ideology presence, intensity, and evolution.
We release two large-scale datasets for noncommercial research use to facilitate further work. Extensive case studies on multiple platforms (X and Truth Social) validate the effectiveness of \framework and provide empirical insights into political polarization and the evolution of online ideologies. Our findings offer a nuanced perspective, advancing both methodological development and empirical understanding in this field.
\end{abstract}



\received{18 March 2024}
\received[revised]{10 October 2025}
\received[accepted]{22 July 2026}

\maketitle

\section{Introduction}
Social media has reshaped political discourse, enabling individuals to express, reinforce, and disseminate ideological views at scale~\cite{zuniga2012social}. A concern in this environment is the formation of echo chambers and the increasing polarization of politics. These effects are particularly visible on platforms where algorithmic curation and social clustering amplify existing biases. In the United States, political opinions are often divided along partisan lines—commonly labeled as ``left'' (Democratic) and ``right'' (Republican). These distinctions influence not only electoral behavior but also patterns of engagement on digital platforms~\cite{pew1,pew2}. This scenario underscores the importance of examining these ideological differences, which have a significant impact on voting behavior and social media activism. In this paper, we aim to contribute a comprehensive methodological framework that not only elucidates these phenomena but also aids in informed policy-making for digital political environments, thereby addressing a vital need in contemporary political discourse.

\begin{figure}[t]
  \centering
  \includegraphics[width=1\textwidth]{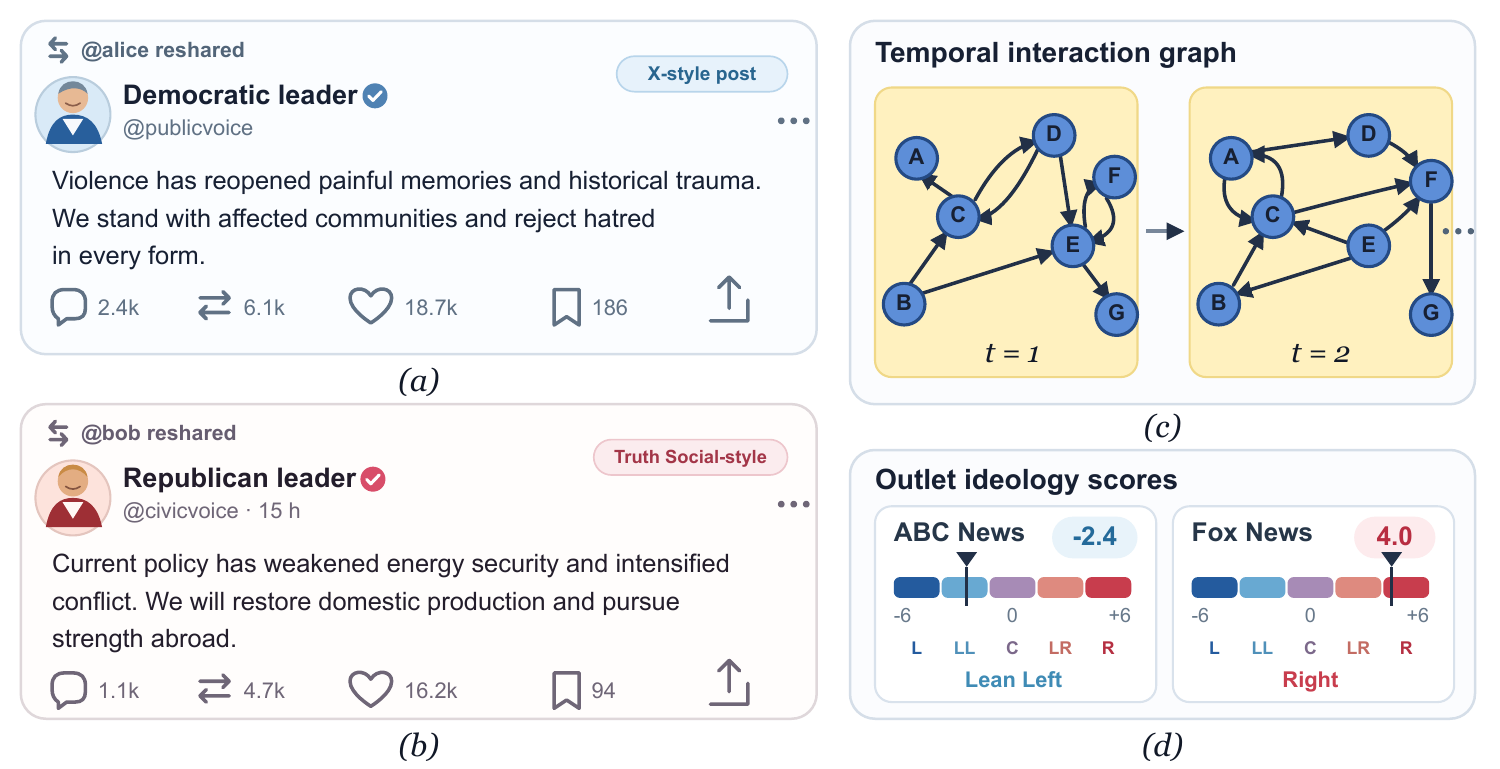}
  \captionsetup{skip=5pt}
  \vspace{-1.1em}
  \caption{Illustration of political ideology dynamics. (a)--(b) Original vector mockups of contrasting X-style and Truth Social-style posts; identities, text, avatars, and engagement counts are illustrative. (c) Temporal graph snapshots. (d) Original five-category scales of AllSides-derived outlet scores.}
  \Description{Four original vector panels illustrate political ideology dynamics: two asset-free mockups of contrasting social posts, directed-network snapshots at time steps 1 and 2, and custom scales placing ABC News at -2.4 (Lean Left) and Fox News at 4.0 (Right). Scale labels L, LL, C, LR, and R denote Left, Lean Left, Center, Lean Right, and Right. No platform logos or profile photographs are used.}
  \label{fig:introduction}
  \vspace{-1.2em}
\end{figure}

The study of political polarization and echo chambers has received increasing attention. Early work investigated ideological clustering on blogs and Facebook~\cite{gilbert2009blogs,quattrociocchi2016echo}, while more recent studies explored polarization during major political events~\cite{flamino2023political}. However, the significance and mechanisms of echo chambers remain debated~\cite{guess2018avoiding}. Yet current research confronts several challenges. First, the issue of data accessibility is prominent; platforms like Twitter impose restrictions on data collection, leading to a scarcity of suitable datasets that align with the size and focus required for comprehensive research. Moreover, the sheer volume of posts on these platforms presents a significant hurdle for manual annotation, a process further complicated by annotator bias. Another challenge is the prevalence of irrelevant content: many social media posts focus on everyday life, lacking clear political signals, and their misclassification can lead to inaccurate predictions of ideological leanings. Effectively filtering this noise to isolate genuine political discourse is paramount for accurate modeling. Additionally, conventional research methods often overlook the dynamic nature of political engagement, missing temporal variations in individual users' political ideologies.
LLMs have shown broad text-processing capabilities~\cite{brown2020language}. Recent work includes retrieval-utility evaluation for retrieval-augmented generation~\cite{dai2025seper}, test-time adaptation~\cite{xu2025you}, cross-domain sequential recommendation~\cite{xin2025llmcdsr}, and multimodal video and spatial reasoning~\cite{guo2025logic,li2025see}. However, direct in-context learning (ICL)~\cite{ouyang2022training} over social media data at scale~\cite{edwards2024language,milios2023context} can incur substantial latency and performance limitations. Watermarking has also been studied for detecting the misuse of open-source LLMs~\cite{xu2025mark}. These constraints motivate a resource-efficient approach tailored to the textual and graph structures of social media.

To address these challenges, we propose a unified framework, \framework (Temporal Social Networks for Political Ideologies), for tracing evolving political ideologies on social media. This framework consists of two core components: the Political Ideology Detection Network (\pidn) and the Political Ideology Prediction Network (\pipn).
\pidn identifies and quantifies the intensity of political ideology in media content. It combines LLM-based style transfer with Unsupervised Domain Adaptation (UDA), aligning structured news (the source) with dynamic, noisy social media (the target). This approach enhances data utility, particularly when annotated social media data is scarce. By focusing on stylistic and semantic cues indicative of political leaning, the \pidn improves the discernment of political ideologies amidst a high volume of irrelevant posts, effectively aiding in filtering non-political content.
Complementing the detection capabilities, \pipn leverages Temporal Graph Neural Networks (TGNNs) to forecast the evolution of users' ideological intensity over time. The TGNN-based \pipn captures the dynamic nature of social media interactions and individual ideological trajectories, enabling prediction of future inclinations. This focus on dynamic, user-level ideological intensity distinguishes our work. While previous long-term analyses typically characterize broader patterns of political communication at the macro level, our approach models the fine-grained, temporal evolution of individual user ideologies. Furthermore, the graph-based nature of our model inherently captures social interaction patterns, offering a more holistic perspective than methods that analyze text in isolation or assume fixed ideological spectrums. Overall, \framework represents a pioneering effort in concurrently determining both the existence and intensity of political ideology among social media users and predicting its evolution.

Our contributions in this paper are multi-fold:
\begin{itemize}
\item We propose \framework, a novel unified framework designed to overcome key challenges in tracing political ideologies on social media. The framework features two integrated modules: 1) The \pidn, which robustly detects ideology by strategically using LLMs for offline style transfer and UDA to bridge the gap between news and noisy social media data, while effectively filtering non-political content. 2) The \pipn, which leverages TGNNs to model and predict the evolution of user ideologies, capturing their presence, intensity, and future trajectory.
\item We contribute two novel, large-scale, publicly released datasets to the research community, significantly addressing the challenge of data accessibility in political discourse analysis on social media platforms. These include: 1) a comprehensive dataset of news sources with curated political ideology scores from AllSides, and 2) a massive, user-centric corpus of nearly 77 million posts from 4,545 X (Twitter) users, spanning 16 years (2007--2022). Both datasets and our code are released for noncommercial research use at \url{https://github.com/yeahjack/TSN4PI}.
\item We conduct extensive experiments on diverse social media platforms (X and Truth Social) to demonstrate the effectiveness and robustness of \framework, validating its capability to analyze complex political discourse and track ideological evolution in real-world settings.
\end{itemize}
\looseness=-1 Our analysis reveals compelling insights into online political behavior. Contrary to the widely accepted echo chamber hypothesis, the users whose ideological positions move the most over time trend predominantly toward the center rather than away from it. This challenges the often-assumed trajectory of increasing polarization over time, highlighting instead a more dynamic and complex landscape of online discourse that warrants empirical, data-driven investigation.

\vspace{-0.35em}
\section{Related Work}

\noindent \textbf{User-centric Political Ideology Detection} has evolved from traditional surveys and manual coding~\cite{achen_1975} to data-driven methods enabled by social media. Early studies used Twitter to infer user leanings~\cite{conover2011}, while \textit{Xiao et al.} introduced TIMME, a graph-based multi-relational embedding method~\cite{xiao2020timme}. Despite progress, prior work often relies on small datasets~\cite{baly-etal-2019-multi}, heuristic or manual annotations, and single data modalities, typically text~\cite{conover2011} or graphs~\cite{xiao2020timme, jiang2023retweet}. Many models also model ideology as binary (left/right) or coarse labels, such as agreement/disagreement~\cite{sapiro2019examining, lo2021}, which limits their ability to capture nuanced stances. 
In contrast, our \framework adopts a granular approach using floating ideological scores, integrating textual and graph-based signals to improve prediction. We acknowledge that political ideology is inherently multidimensional, beyond a single left-right axis. For example, \textit{Colacrai et al.}~\cite{colacrai2024navigating} analyzed Reddit's political compass, disentangling ideology into economic (left-right) and social (libertarian-authoritarian) axes. Inspired by this, we adopt the widely-studied one-dimensional continuum as a tractable starting point for modeling polarization, while recognizing multidimensional extensions as a promising direction for future work. Our framework further supports expansion to issue-level or temporal perspectives on behavior.

\noindent \textbf{Opinion Evolution and Political Dynamics on Social Media} involve phenomena such as echo chambers (where users are exposed predominantly to ideologically aligned content) and political polarization, often reinforced by platform algorithms. These dynamics are central to understanding how public opinion forms and shifts in online environments.  
For example, \textit{Chen et al.}~\cite{chen2024public} proposed a graph-clustering method to model the evolution of public opinion topics on social media during Pelosi's visit to Taiwan, revealing causal transitions between topics across event stages. Other studies focused on long-term communication patterns of political figures. \textit{Oliveira et al.}~\cite{oliveira2021longterm} analyzed six years of Brazilian politicians’ communications, addressing concept drift by designing classifiers to distinguish political from non-political posts. Their findings showed that communication strategies and public reactions varied around major events~\cite{oliveira2021longterm}.
While prior work offers insights into topic-level discourse and macro trends, our approach differs in both focus and granularity. Instead of classifying content types or tracking broad topic shifts, we model continuous, user-level ideological scores over time, a task methodologically related to trajectory prediction~\cite{wang2021variable} and user modeling from implicit feedback~\cite{wang2018confidence, wang2020setrank, wang2021personalized}. By combining semantic signals with dynamic interaction patterns, our framework enables fine-grained tracking of individual ideological evolution.

\noindent \textbf{Semantic Text Analysis for Political Ideology} has advanced rapidly with NLP developments. Early work employed RNNs for partisan classification~\cite{iyyer2014political}, laying the groundwork for computational ideology detection. The advent of transformer-based models like BERT~\cite{devlin2018bert} enabled more precise text representations and improved classification performance.
Later variants—\texttt{RoBERTa}~\cite{liu2019roberta}, \texttt{SBERT}~\cite{reimers2019sentence}, \texttt{BERTweet}~\cite{nguyen2020bertweet}, and \texttt{TWHIN-BERT}~\cite{zhang2023twhin}—further enhanced effectiveness on social media data. More recently, LLMs like GPT-3~\cite{brown2020language} have enabled zero-/few-shot political text understanding via in-context learning. However, their direct use for large-scale, fine-grained prediction remains limited by inference cost and stability.
Our framework adopts a modular design, using an LLM not as an end-to-end predictor but as a style transfer module. This bridges the domain gap between structured political news and noisy social media posts, as detailed in Section~\ref{sec:methods}.

\noindent \textbf{Stance Detection} is a related task in NLP that involves identifying an expressed position—such as ``favor'', ``against'', or ``neutral''—towards a specific target~\cite{mohammad2016semeval,kucuk2020stance,sobhanifebruary2017dataset}.
While both stance detection and political ideology detection analyze opinions, they differ in scope. Stance is target-specific and often transient, reflecting an individual’s position on a particular issue~\cite{augenstein2016stance,kucuk2020stance}. In contrast, political ideology is broader and more stable, representing a cross-topic belief system that shapes multiple stances~\cite{baly-etal-2019-multi}. Our work focuses on identifying and modeling this fundamental ideological attribute over time, rather than classifying positions on specific targets.

\vspace{-0.35em}
\section{Data}

\begin{figure}[t]
    \centering
    \includegraphics[width=0.8\linewidth]{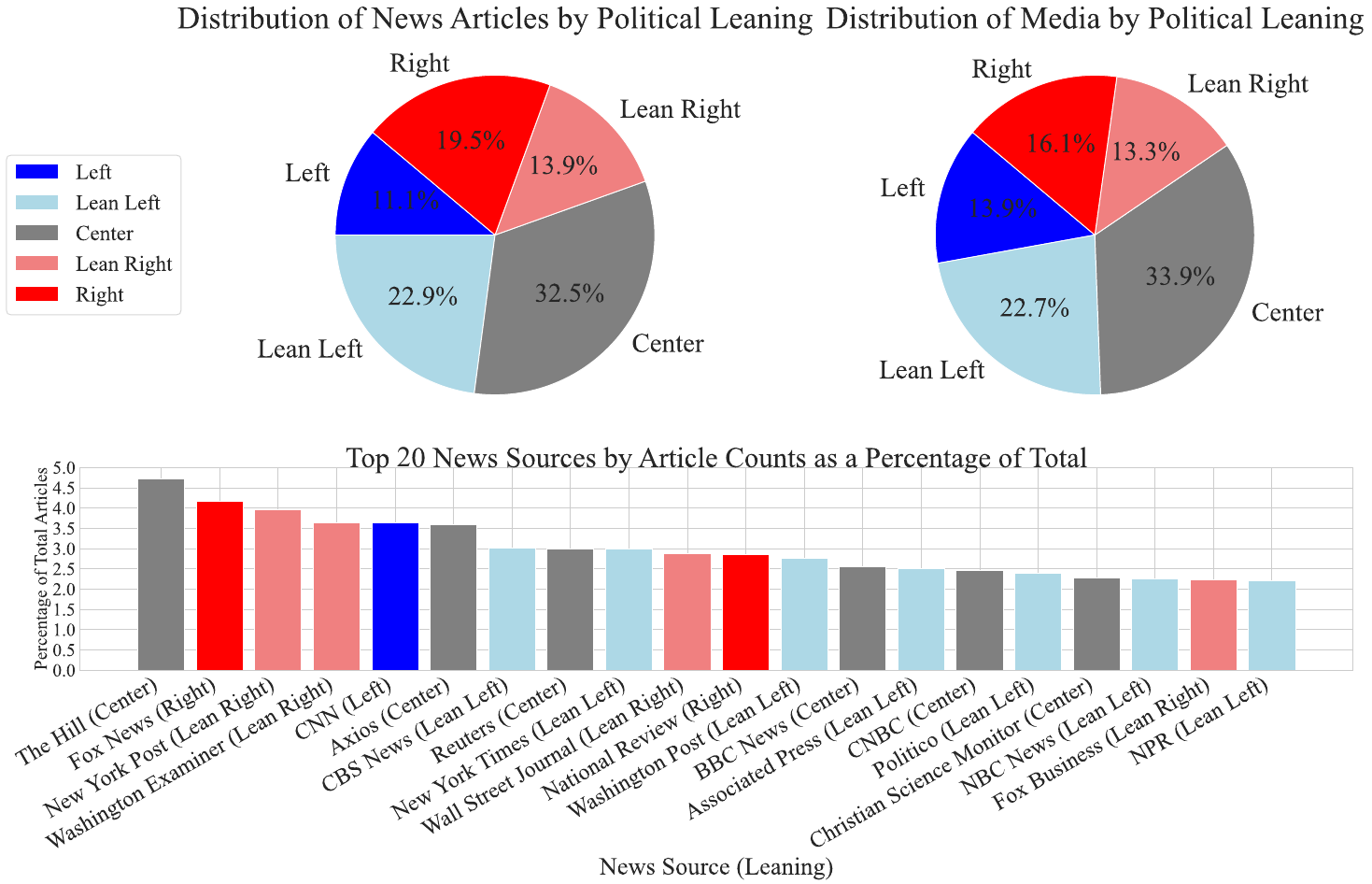}
    \vspace{-0.25em}
    \caption{Distribution of articles (left) and outlets (right) by political leaning, based on data from AllSides. The bar graph shows the top 20 sources by share of total articles, colored by ideology.}
    \Description{Two pie charts show news-article and media-outlet shares across five political leanings, with Center the largest category in both. A bar chart below ranks the top 20 outlets by article share, with The Hill highest.}
    \label{fig:distribution_allsides}
\end{figure}

We use three datasets in this study. The first, from \url{allsides.com}, provides ideological labels for news sources and supports cross-domain adaptation. The second and third contain social media posts from X (formerly Twitter) and Truth Social, respectively. These platforms are used for model training, evaluation, and cross-platform analysis of political discourse and user ideology across distinct online communities. Privacy considerations are detailed in \smref{Appendix~A.1.1}.\footnote{All appendices are provided in the supplementary material submitted alongside this article.}

\vspace{-0.35em}
\subsection{News Political Ideology Dataset}

We construct a political ideology dataset using media bias ratings from \url{allsides.com}~\cite{allsides_ratings}, which aggregates partisan perspectives across the ideological spectrum.\footnote{AllSides Media Bias Ratings\texttrademark{} by AllSides.com, used with permission under CC~BY-NC~4.0.} This dataset supports modeling and transferring ideological signals across domains, based on the assumption that media affiliation reflects consistent leanings. \url{allsides.com} applies a multi-partisan methodology, including editorial review, blind bias surveys, third-party analysis, and community feedback. Each news source is assigned a continuous bias score from -6 (most left) to 6 (most right), grouped into five categories: \textit{Left} ($[-6, -3]$), \textit{Lean Left} ($(-3, -1]$), \textit{Center} ($(-1, 1]$), \textit{Lean Right} ($(1, 3]$), and \textit{Right} ($(3, 6]$). Using \texttt{requests} and \texttt{Newspaper3k}~\cite{newspaper3k}, we retrieved rated outlets and crawled $\sim$230,000 articles from their main pages, covering August 3, 2017 to December 22, 2022. Each article inherits its source's label, yielding a large-scale, weakly supervised dataset with balanced ideological coverage. Score preprocessing details are in \smref{Appendix~A.1.2}.

\subsubsection*{\textbf{Preliminary Data Analysis.}}
We conducted a preliminary analysis of the AllSides dataset to examine its ideological distribution and linguistic characteristics. The upper part of Figure~\ref{fig:distribution_allsides} shows the proportions of news articles and media outlets across five political categories. The dataset displays a relatively balanced Left-Right distribution, supporting class balance during model training. The lower part of Figure~\ref{fig:distribution_allsides} illustrates the share of total articles held by the top 20 news outlets. Among them, ``The Hill,'' labeled as Center, contributes the largest share. Despite a Right-leaning tilt among the very largest outlets, these 20 sources span all five ideological categories. Additional distribution details and word cloud visualizations appear in \smref{Table~1 and Figure~1} of that material.

Beyond word cloud visualizations, we conducted a frequency analysis of full article texts to examine ideological differences in lexical usage. Table~\ref{tab:word-frequency-by-leaning} lists the top 10 most frequent content words per political leaning after stopword removal. These patterns reveal how lexical emphasis varies across ideological contexts.
\textit{Trump} is the most frequent term in both Left (96{,}727) and Right (60{,}993) articles, indicating shared focus but likely divergent framing. \textit{Biden} also appears frequently in Lean Right (45{,}403) and Right (50{,}717), reflecting sustained—often critical—attention from right-leaning outlets.
An asymmetry emerges with \textit{covid}, ranked 6\textsuperscript{th} in Lean Left (86{,}424) but absent elsewhere, suggesting greater pandemic focus among moderately left-leaning sources.
Center-leaning content foregrounds governing bodies—\textit{state} (122{,}494), \textit{house} (95{,}068), \textit{senate} (82{,}457)—indicating a governance-oriented tone. Lean Right and Right texts instead emphasize oversight and platform arenas, with \textit{congress} (34{,}186), \textit{committee} (25{,}647), and \textit{twitter} (34{,}956), reflecting political critique and media discourse.
These trends support our hypothesis that bias emerges not only in topic selection, but also in lexical framing—even when events are shared.

\begin{table}[t]
\centering
\caption{Top 10 most frequent content words per political leaning (full text, stopwords removed).}
\label{tab:word-frequency-by-leaning}
\rowcolors{2}{gray!10}{white} %
\resizebox{\linewidth}{!}{
\begin{tabular}{@{}cccccc@{}}
\toprule
\textbf{Rank} & \textbf{Left} & \textbf{Lean Left} & \textbf{Center} & \textbf{Lean Right} & \textbf{Right} \\
\midrule
1  & trump (96,727)      & people (130,164)     & state (122,494)     & president (54,261)  & trump (60,993) \\
2  & people (62,773)     & president (96,049)   & trump (121,089)     & biden (45,403)      & biden (50,717) \\
3  & election (58,163)   & state (94,501)       & president (115,675) & house (42,055)      & house (50,398) \\
4  & republican (47,839) & new (92,173)         & people (109,190)    & new (41,471)        & president (47,401) \\
5  & even (47,647)       & trump (88,297)       & new (105,062)       & state (40,075)      & news (47,398) \\
6  & president (46,673)  & covid (86,424)       & house (95,068)      & congress (34,186)   & people (40,715) \\
7  & political (43,828)  & biden (79,017)       & year (84,825)       & year (28,292)       & new (37,368) \\
8  & house (42,453)      & years (72,749)       & states (83,646)     & former (28,068)     & state (35,696) \\
9  & democrats (41,652)  & states (72,083)      & senate (82,457)     & last (26,605)       & twitter (34,956) \\
10 & republicans (38,995)& election (65,009)    & election (82,358)   & committee (25,647)  & government (33,186) \\
\bottomrule
\end{tabular}
}
\vspace{-0.3em}
\end{table}

\vspace{-0.35em}
\subsection{A User-Centric Twitter Dataset on U.S. Presidential Elections}
\begin{wraptable}{r}{0.3\linewidth}
  \centering
  \footnotesize
  \vspace{-1.0em}
  \caption{X Dataset Overview.}
  \label{tab:twitter_overview_combined}
  \rowcolors{2}{gray!10}{white} %
  \subcaptionbox{Basic Statistics\label{tab:twitter_overview}}[0.95\linewidth]{%
    \vspace{-0.75em}
    \centering
    \begin{tabular}{@{}lr@{}}
      \toprule
      \textbf{Characteristic} & \textbf{Value} \\
      \midrule
      Number of Users       & 4,545 \\
      Total Tweets          & 76,718,924 \\
      Avg. Tweets/User      & 16,880 \\
      Avg. Retweets/Tweet   & 20.50 \\
      Avg. Likes/Tweet      & 88.00 \\
      Avg. Quotes/Tweet     & 2.40 \\
      \bottomrule
    \end{tabular}
  }

  {%
    \centering
    \subcaptionbox{Attributes\label{tab:twitter_attributes}}{%
    \vspace{-2em}
      \resizebox{1.0\linewidth}{!}{%
        \begin{tabular}{@{}llll@{}}
          \toprule
          \textbf{Field} & \textbf{Relations} & \textbf{Target ID} & \textbf{Content} \\
          \midrule
          ID        & Retweets   & RT~ID      & Text \\
          Username  & Quotes     & QT~ID      & Hashtags \\
          Time      & Likes      & Reply~ID   & \\
          Conv.\,ID &            & Reply User & Mentions \\
          \bottomrule
        \end{tabular}
      }
    }
    \par
  }%
\end{wraptable}
To support research on political ideology and discourse—particularly in the context of U.S. presidential elections—we construct a large-scale dataset from X (formerly Twitter). Using a keyword seed query (``presidential election''), we collected $\sim$10,000 tweets and identified 4,545 unique users. We then retrieved their full tweet histories, yielding a user-centric corpus of 76,718,924 tweets from January 1, 2007 to December 31, 2022 (Table~\ref{tab:twitter_overview}). This dataset enables longitudinal analysis of political discourse and user behavior.
Table~\ref{tab:twitter_overview} summarizes dataset statistics, and Table~\ref{tab:twitter_attributes} lists tweet-level fields: tweet IDs, user metadata, timestamps, conversation IDs, engagement metrics (retweets, quotes, likes), reference links (e.g., replies, retweets), and full text with hashtags. The dataset’s feature richness supports detailed analysis of user activity, engagement patterns, and content evolution. Preprocessing details appear in \smref{Appendix~A.1.3}.

\subsubsection*{\textbf{Comparison with Existing Datasets.}}

Table~\ref{tab:social_media_datasets_summary} compares recent large-scale social media datasets. While many are valuable, they vary in platform, scale, granularity, and thematic scope. For example, Dreaddit~\cite{turcan2019dreaddit} and the Reddit Corpus~\cite{henderson2019repository} focus on Reddit, providing post- or dialogue-level data (190k and 727M entries, respectively). Weibo-COV~\cite{hu2020weibo} offers 40M COVID-19-related posts from Sina Weibo, but is also post-centric. Koo~\cite{mekacher2024koo} (72M posts from 1.4M users) and iDRAMA-Scored~\cite{patel2024idrama} (57M posts from 226k users) target Indian and fringe-platform discourse, respectively, and are likewise post-level. Even Twitter-based datasets such as TweetIntent@Crisis~\cite{ai2024tweetintent} and News Feed Ranking~\cite{boukhalfa2022ranking} involve small user bases (e.g., 79 and 46 users) and are not user-centric.

In contrast, our dataset offers full tweet histories for a fixed cohort of 4,545 users over 16 years. This supports fine-grained, longitudinal modeling of U.S. presidential election discourse. Unlike post-level snapshots, our dataset retains full threads and engagement, allowing for the analysis of ideological evolution and stance shifts over time. Its scale, temporal span, and user continuity make it uniquely suited for studying political polarization and ideology formation on social media.

\begin{table}[tbp]
  \caption{Summary of Recent Social Media Datasets}
  \label{tab:social_media_datasets_summary} %
  \centering %
  \footnotesize
  \setlength{\tabcolsep}{3pt}
  \rowcolors{2}{gray!10}{white} %

  \begin{tabularx}{\linewidth}{%
    >{\centering\arraybackslash}p{2.9cm}  %
    >{\centering\arraybackslash}p{1.6cm}  %
    >{\centering\arraybackslash}p{1.2cm}  %
    >{\centering\arraybackslash}p{0.8cm}  %
    >{\centering\arraybackslash}p{1.5cm}  %
    >{\centering\arraybackslash}p{1.4cm}  %
    >{\centering\arraybackslash}X         %
  }
    \toprule
    \textbf{Dataset Name} & \textbf{Platform} & \textbf{Count} & \textbf{Users} & \textbf{Duration} & \textbf{Data Level} & \textbf{Theme} \\
    \midrule

    Dreaddit~\cite{turcan2019dreaddit} & Reddit & $\sim$190k & N/A & 2017--2018 & Post & Social media stress \\

    Reddit Corpus~\cite{henderson2019repository} & Reddit & 727M & N/A & 2015--2018 & Dialogue & Conversational AI \\

    Weibo-COV~\cite{hu2020weibo} & Sina Weibo & >40M & 20M & 2019--2020 & Post & COVID-19 discourse \\

    News Feed Rank.~\cite{boukhalfa2022ranking} & Twitter & 26k & 46 & 2021--2022 & News & Personalized ranking \\

    MetaHate~\cite{piot2024metahate} & Mixed & 1.2M & N/A & N/A & Post & Hate speech detect. \\

    IsamasRed~\cite{chen2024isamasred} & Reddit & 8M & N/A & 2023 & Dialogue & Israel-Hamas discourse \\

    Koo Dataset~\cite{mekacher2024koo} & Koo & 72M & 1.4M & 2020--2023 & Post & Indian microblogging \\

    TweetIntent Crisis~\cite{ai2024tweetintent} & Twitter & $\sim$18k & 79 & 2022--2023 & Post & Russia-Ukraine narratives \\

    SocialDrought~\cite{shang2024socialdrought} & Twitter\&News & 1.5M & N/A & 2012--2023 & Integrated & Drought societal impact \\

    LGBTQ+ MiSSoM~\cite{cascalheira2024missom} & Reddit & 27.7k & N/A & 2012--2021 & Post & LGBTQ+ stress research \\

    AiGen-FoodRev.~\cite{gambetti2024aigen} & Yelp & 20k & N/A & N/A & Post & AIGC Detection \\

    iDRAMA-Scored~\cite{patel2024idrama} & Scored & 57M & 226k & 2020--2023 & Post & Fringe communities \\

    Unintended Offense~\cite{tsai2024leveraging} & Twitter & 2.4k & N/A & N/A & Dialogue & Unintentional offense \\

    PulseReddit~\cite{han2025pulsereddit} & Reddit & $\sim$73k & N/A & 2024--2025 & Post & MAS HFT Crypto Trad. \\
    \midrule %
    \textbf{Ours} & Twitter & \textbf{$\sim$77M} & \textbf{$\sim$5k} & \textbf{2007--2022} & \textbf{User} & \textbf{Presidential Election} \\

    \bottomrule
  \end{tabularx}
  \vspace{-1.0em}
\end{table}

\subsubsection*{\textbf{Preliminary Data Analysis.}}

To better understand the dataset’s temporal growth and topical focus, we carried out exploratory analyses of tweet volume over time and hashtag frequency.

\begin{figure}[h]
  \centering
  \vspace{-0.5em}
  \begin{subfigure}[t]{0.48\textwidth}
    \centering
    \includegraphics[width=\linewidth]{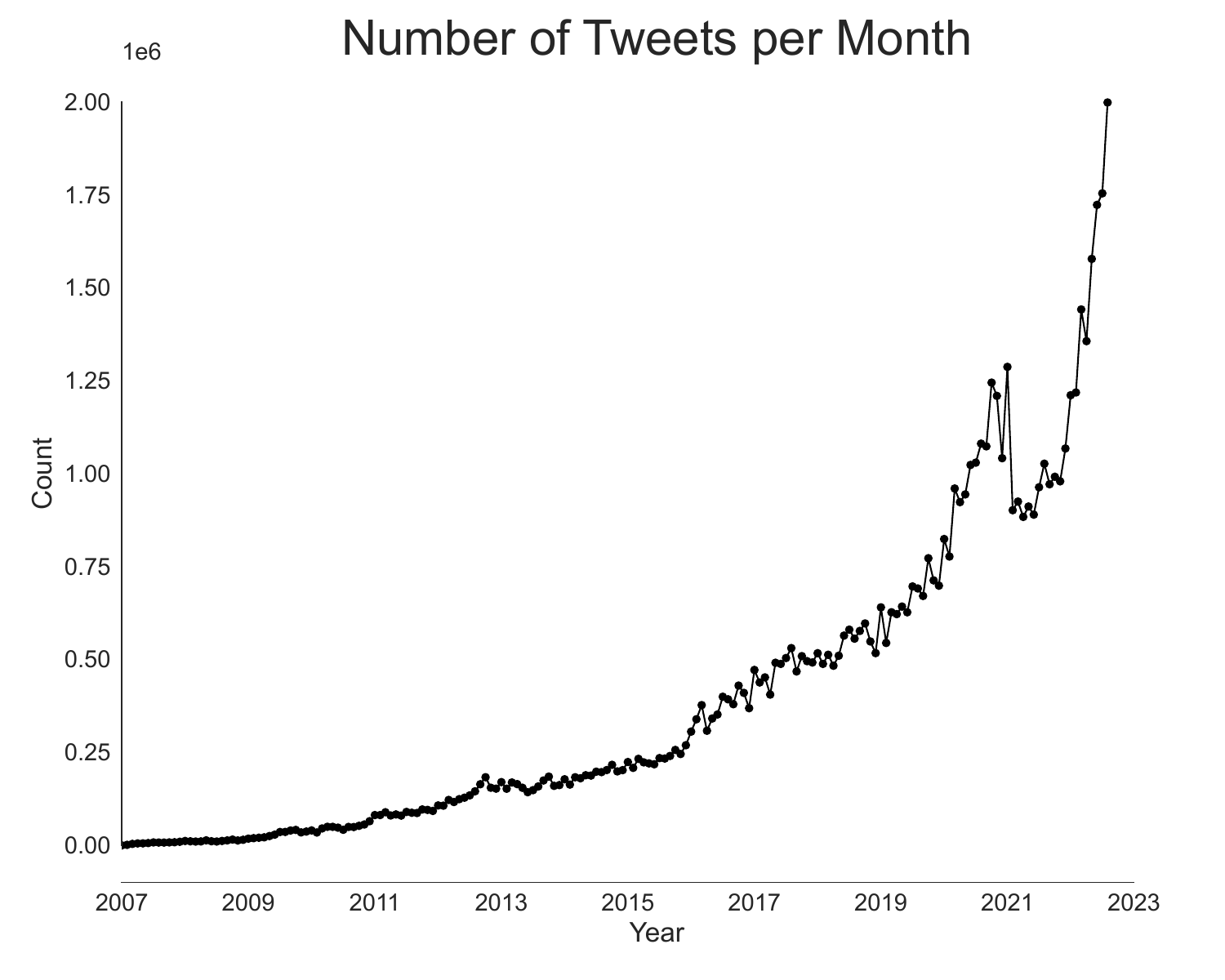}
    \caption{Number of tweets per month from January~2007 to September~2022.}
    \label{fig:tweets_per_month}
  \end{subfigure}
  \hfill
  \begin{subfigure}[t]{0.48\textwidth}
    \centering
    \includegraphics[width=\linewidth]{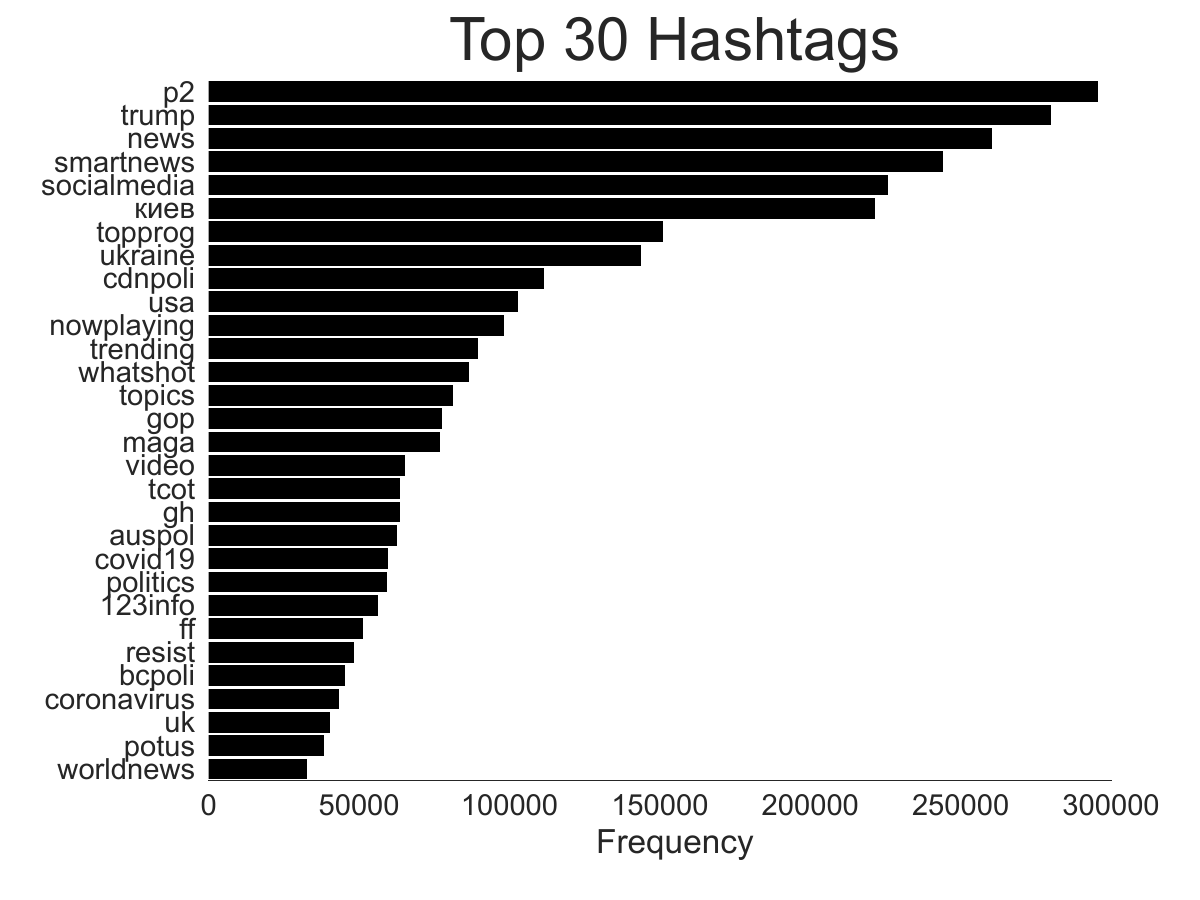}
    \caption{Top~30 hashtags in the Twitter dataset by frequency.}
    \label{fig:top_hashtags}
  \end{subfigure}
  \captionsetup{skip=4pt}
  \caption{Temporal growth and topical focus of the X (Twitter) dataset.}
  \Description{Two panels summarize the X dataset. The left line chart shows monthly tweet counts rising from near zero in 2007 to roughly two million by late 2022, with intermediate spikes. The right horizontal bar chart ranks 30 hashtags, led by p2, trump, news, smartnews, and socialmedia.}
  \label{fig:combined_twitter_data}
\end{figure}

\noindent\textbf{Tweet Volume Over Time.}
Figure~\ref{fig:tweets_per_month} shows monthly tweet counts from Jan.~2007 to Sep.~2022. The overall trajectory is upward, with tweet activity growing steadily through late 2022 and accelerating especially rapidly after 2018. Four peaks punctuate this trend: a modest bump during the 2012 U.S. presidential election, a more pronounced surge in 2016, a third spike during the COVID-19 outbreak in early 2020, and the largest peak surrounding the 2020 presidential election, reflecting heightened online political activity. Afterward, tweet volume continues to rise sharply, suggesting a sustained expansion of political engagement on the platform.

\noindent\textbf{Top Hashtags.}
Figure~\ref{fig:top_hashtags} lists the 30 most frequent hashtags. News- and media-oriented tags, such as \texttt{\#news}, \texttt{\#smartnews}, and \texttt{\#socialmedia}, appear most prominently, underscoring Twitter’s role as a live information channel. Excluding these politically neutral news tags, partisan motifs—\texttt{\#gop}, \texttt{\#maga}, and \texttt{\#resist}—still command a substantial share of the remaining distribution, confirming the platform’s ideological polarity. International-politics hashtags (\texttt{\#ukraine}, \texttt{\#auspol}, \texttt{\#cdnpoli}) and global-health hashtags (\texttt{\#covid19}, \texttt{\#coronavirus}) are likewise present, highlighting the dataset’s geographic breadth and its coverage of worldwide events.

\vspace{-0.35em}
\subsection{Truth Social Dataset}

We also incorporated a dataset from Truth Social~\cite{gerard2023truth}, launched in February 2022 following the January 2021 suspension of then-U.S. President Donald Trump from multiple platforms, including Twitter (now known as X), Facebook, and others. From Figure~\ref{fig:introduction} (a) and (b), Truth Social was designed to closely mirror X in terms of functionality, with almost identical features. For instance, X's ``Retweet'' functionality is analogous to ``ReTruth'' on Truth Social. Supporters of Trump predominantly frequent this platform and generally exhibit a right-leaning political bias. With this dataset, we aim to explore the dissemination of information and track the evolution of user political ideologies within a social network that inherently carries a political bias from the outset.

\begin{wraptable}{r}{0.25\linewidth}
\centering
\vspace{-0.25em}
\small
\vspace{-0.9em}
\caption{Truth Social Dataset Overview}
\label{tab:truth_social_dataset_overview}
\rowcolors{2}{gray!10}{white} %
\begin{tabular}{lr}
\toprule
\textbf{Field} & \textbf{\# of Records} \\
\midrule
Users   & 454,458 \\
Truths  & 823,927 \\
Quotes  & 10,508 \\
Replies & 506,276 \\
\bottomrule
\end{tabular}
\vspace{-1.0em}
\end{wraptable}

Structurally, Truth Social bears similarities to Twitter. In contrast to Twitter's ``Tweets,'' posts on this platform are referred to as ``Truths,'' and reposts are termed ``ReTruths.'' The dataset comprises over 454,000 users and exceeds 823,000 ``Truths.'' Additionally, it includes the complete post history of the 65,536 most active users. Subtables within this dataset, such as ``Truths'' and ``Users,'' serve as primary resources for our exploration. An overview of this dataset is presented in Table~\ref{tab:truth_social_dataset_overview}. Details of data preprocessing are provided in \smref{Appendix~A.1.4}.

\vspace{-0.35em}
\section{Methods}
\label{sec:methods}
This section introduces our proposed framework and consists of both \pidn and \pipn modules.

\begin{figure*}[t]
  \centering
  \includegraphics[width=1.0\textwidth]{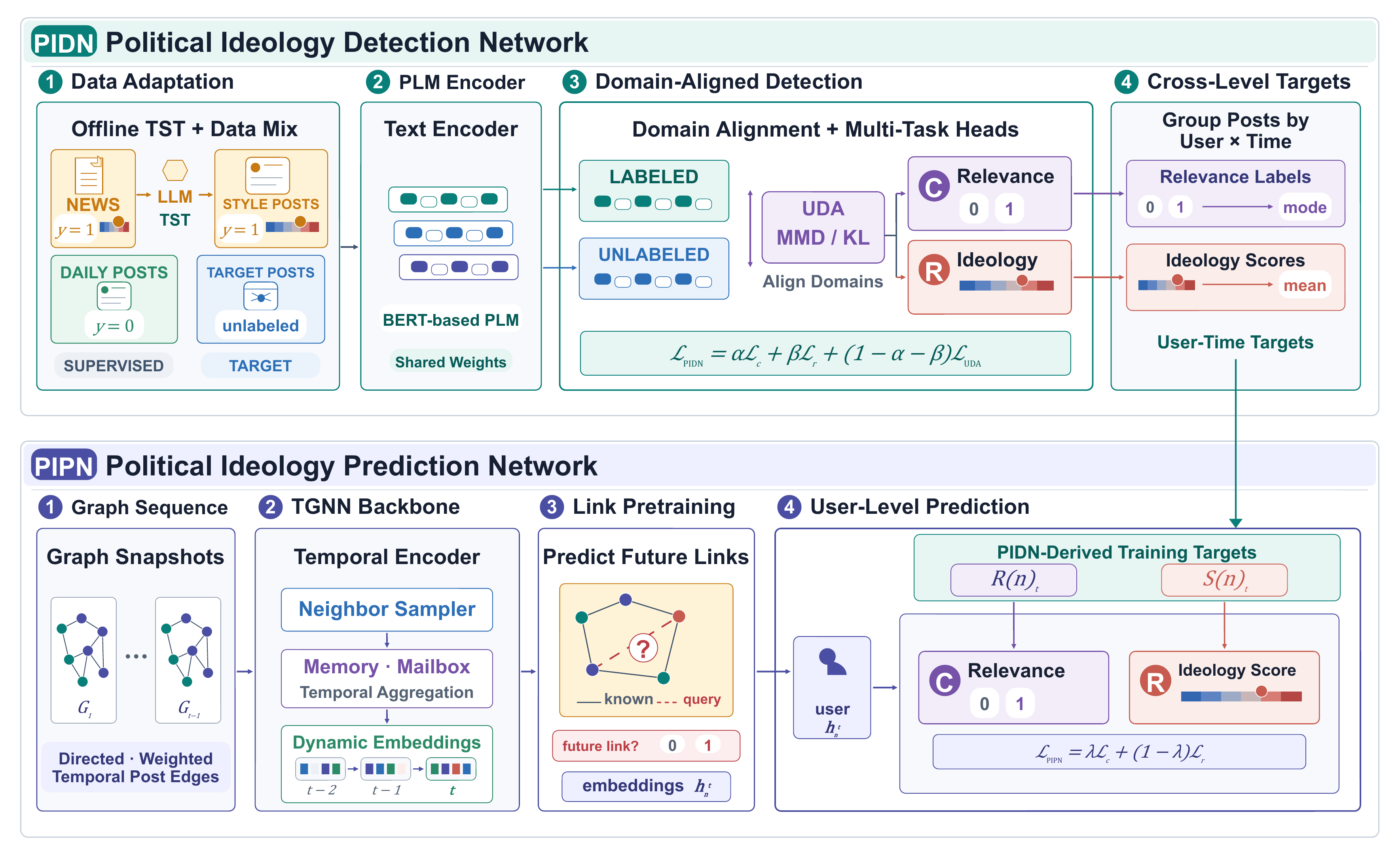}
  \vspace{-1.25em}
  \caption{The overview of our proposed framework: \framework. The upper part of the figure represents the \pidn module, while the lower part represents the \pipn module.}
  \Description{A two-stage pipeline. The upper PIDN transforms news into social-media style, combines it with daily-life and unlabeled social-media posts, encodes them with a BERT-based language model, and optimizes classification, regression, and domain-alignment losses. The lower PIPN processes temporal social graphs with a temporal graph neural network for link prediction and node classification and regression.}
  \label{fig:framework}
\end{figure*}

\vspace{-0.35em}
\subsection{Political Ideology Detection Network}
Political ideology detection involves analyzing social media posts to determine 1) whether they are politically relevant and, if so, 2) to what extent they align along the political spectrum. Rather than assigning scores to all inputs indiscriminately, the task first filters out non-political content and then estimates an ideology score for posts identified as political. Our proposed network addresses these dual challenges through a combination of unsupervised domain adaptation (UDA), text style transfer (TST) using LLMs, and an efficient model architecture.

While LLMs show strong ICL capabilities~\cite{brown2020language,dong2023survey}, their direct use in large-scale tasks like ideology detection faces high inference latency. Adding in-context examples lengthens the input and increases attention computation~\cite{zhao2023survey,pope2023efficiently,wu-etal-2025-tokenselect}. LLMs can also underperform fine-tuned models on well-defined, high-resource tasks~\cite{perez-etal-2025-exploring}, limiting their suitability for real-time or high-throughput use.
We therefore adopt a BERT-based architecture, which offers a stronger trade-off between task performance and efficiency—especially in inference speed and resource usage~\cite{sanh2019distilbert}. LLMs are reserved for the offline TST stage, where latency is less critical, rather than the core detection pipeline.

\setlength{\columnsep}{4pt}
\begin{wrapfigure}[14]{r}{0.48\linewidth}
    \centering
    \vspace{-0.5em}
    \includegraphics[width=\linewidth]{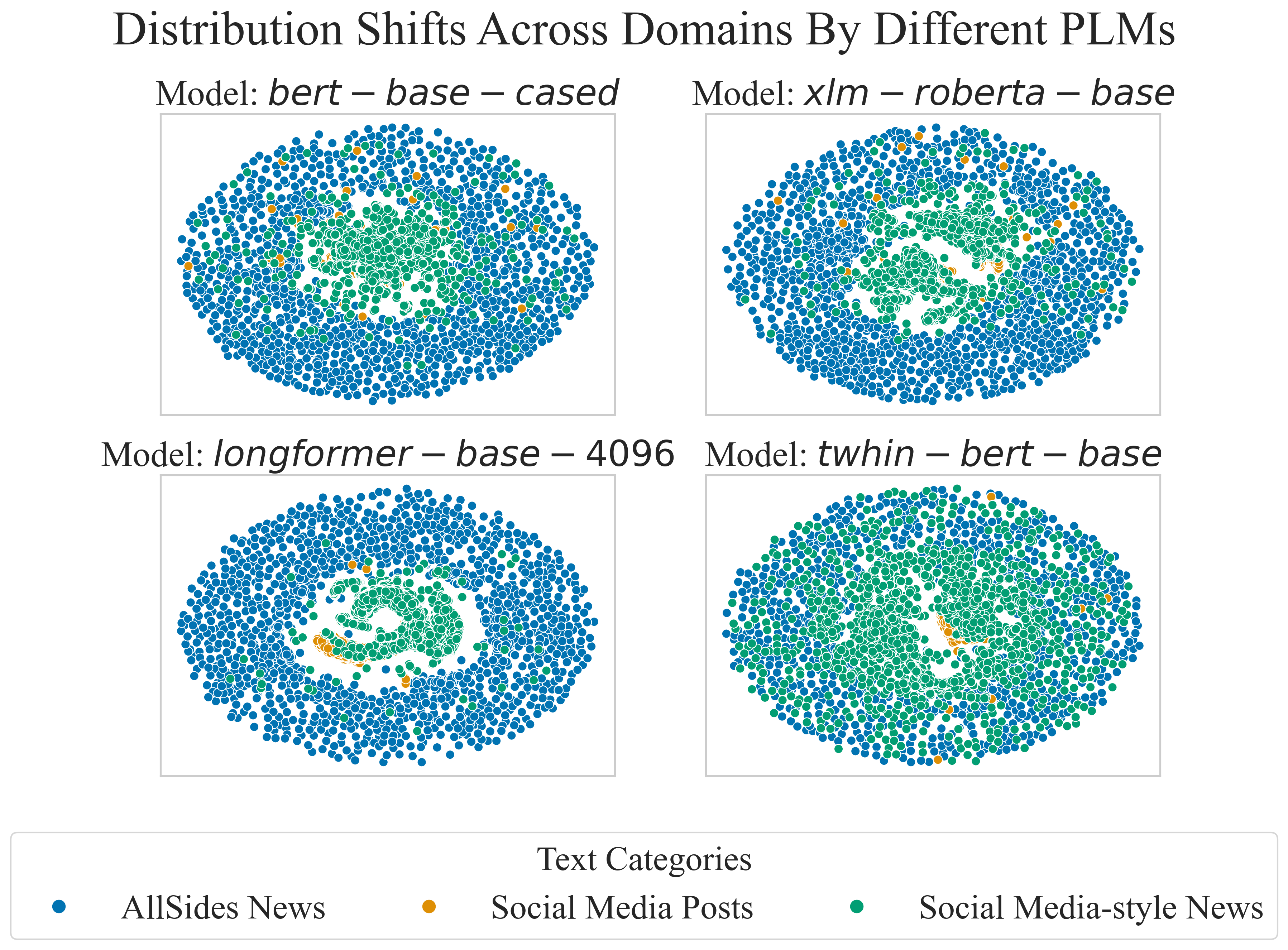}  %
    \caption{Embedding across PLMs.}
    \Description{Four UMAP scatter plots compare AllSides news, social-media posts, and style-transferred news under BERT, XLM-R, Longformer, and TWHIN-BERT embeddings. Across models, the style-transferred news shifts toward the social-media cluster.}
    \label{fig:distribution_shifts}
\end{wrapfigure}
Building on this architectural decision, our detection pipeline integrates offline LLM-based style adaptation with an online BERT-based classifier. Specifically, given the substantial stylistic differences between formal news articles and informal social media posts, we apply a TST module to transform news content into a style resembling that of platforms such as X. This stylistic alignment narrows the domain gap between the labeled source domain (news) and the unlabeled target domain (social media), thereby improving transferability.
To further bridge this gap, we incorporate UDA to align the latent representations of political news and unlabeled social media posts. We also augment training with a mixture of political news articles and non-political social media content, allowing the model to jointly learn 1) political relevance classification and 2) ideology scoring for relevant posts. This unified approach enables robust detection in noisy, real-world social media environments, while leveraging structured ideological signals from traditional media.

\subsubsection*{\textbf{LLMs for Text Style Transfer.}}

To bridge the stylistic gap between formal news articles and the colloquial nature of social media posts, we use LLMs to perform TST. This approach is especially suitable given the absence of parallel corpora and the strong instruction-following capabilities of modern LLMs. Specifically, for each news article \( n \) in the source corpus \( N \), we generate a style-aligned version \( n_s \) by prompting the LLM with: $n_s = \text{LLM}([\text{prompt}, n]), n \in N.$
\label{method_llm_tst}
To intuitively validate our approach, we sampled texts from three groups: the source domain (AllSides news), the target domain (Truth Social posts), and our generated TST outputs. These were encoded using four diverse BERT-based PLMs: \texttt{bert-base-cased}~\cite{devlin2018bert}, \texttt{xlm\allowbreak -roberta\allowbreak -base}~\cite{liu2019roberta}, \texttt{longformer\allowbreak -base\allowbreak -4096}~\cite{beltagy2020longformer}, and \texttt{twhin\allowbreak -bert\allowbreak -base}~\cite{zhang2023twhin}—and projected to 2D via UMAP~\cite{SMG2020}.
\looseness=-1 As shown in Figure~\ref{fig:distribution_shifts}, original news articles (blue) are widely scattered, while social media posts (orange) form a denser cluster. Crucially, style-transferred texts (green) align closely with the target domain, reducing the distributional gap. This pattern holds across all four PLMs, supporting the effectiveness and robustness of our TST strategy. A quantitative and comprehensive automatic evaluation follows in Section~\ref{sec:tst_evaluation}.

\subsubsection*{\textbf{Unsupervised Domain Adaptation for Distribution Alignment.}}
While LLM-based TST narrows the stylistic gap between news and social media texts, residual distributional shifts remain. To address this, particularly without requiring labeled social media data, we incorporate an optional UDA module designed to enhance domain generalization.
Our primary approach employs Maximum Mean Discrepancy (\textsc{MMD}), a non-parametric measure that quantifies the divergence between two distributions in a reproducing kernel Hilbert space (RKHS), without requiring paired or balanced samples. Given a text sample \( T \), we extract its embedding \( E = L(T) \) using a PLM \( L \). Let \( E_X \) and \( E_Y \) denote embeddings from the style-transferred source and target domains, respectively. The \textsc{MMD} loss is computed as:
\begin{equation}
\mathcal{L}_{\textsc{MMD}}(E_X, E_Y) = \sqrt{\mathbb{E}[k(x,x')] + \mathbb{E}[k(y,y')] - 2\mathbb{E}[k(x,y)]},
\end{equation}
where \( k(x, y) = \exp\left(-\|x - y\|^2/\left(2\sigma^2\right)\right) \) is the Gaussian RBF kernel, and \( x, x' \in E_X \), \( y, y' \in E_Y \). This kernel maps samples into a space where their similarity can be effectively measured and optimized.

In addition to \textsc{MMD}, we also consider Kullback-Leibler (\textsc{KL}) divergence as an alternative distribution alignment method. \textsc{KL} divergence is particularly effective when model outputs can be interpreted as probability distributions—such as softmax scores or normalized embeddings—allowing explicit alignment of posterior distributions between domains. Given source domain prediction \( P \) and target domain prediction \( Q \), the \textsc{KL} loss is computed as:
\begin{equation}
\mathcal{L}_{\textsc{KL}}(P \,\|\, Q) = \sum_i P(i) \log \frac{P(i)}{Q(i)}.
\end{equation}
Compared to \textsc{MMD}, \textsc{KL} divergence assumes stronger alignment in probability space and is more suitable when output distributions are well-calibrated. We evaluate both metrics to assess robustness under varying domain conditions. To incorporate domain adaptation during training, we define a composite loss that jointly optimizes task performance and representation alignment:
\begin{equation}
\mathcal{L}_{\text{PIDN}} = \alpha \mathcal{L}_c + \beta \mathcal{L}_r + (1 - \alpha - \beta)\mathcal{L}_{\text{UDA}},
\end{equation}
where \( \mathcal{L}_c \) and \( \mathcal{L}_r \) are the political-relevance classification and ideology-regression losses, respectively; \( \mathcal{L}_{\text{UDA}} \in \{\mathcal{L}_{\textsc{MMD}}, \mathcal{L}_{\textsc{KL}}\} \) is the selected domain-alignment loss; and \( \alpha,\beta \ge 0 \) with \( \alpha+\beta\le 1 \) control their relative contributions. When UDA is disabled in an ablation, the \( \mathcal{L}_{\text{UDA}} \) term is omitted.
By aligning embedding distributions across domains, the UDA module enhances generalization to informal, unlabeled social media content while preserving task-specific learning. A comprehensive complexity analysis of the UDA module is provided in \smref{Appendix~A.4}.

\vspace{-0.35em}
\subsection{Political Ideology Prediction Network}
Political ideology prediction models how users’ stances evolve over time—a task fundamentally different from static detection due to the absence of future posts. Traditional post-level embedding methods are thus inapplicable. We adopt Temporal Graph Neural Networks (TGNNs), which are well-suited for ideological forecasting. This choice is motivated by two factors: social media platforms naturally produce dynamic user interaction graphs, and TGNNs are designed to model temporal dependencies in such evolving structures.

\subsubsection*{\textbf{Temporal Graph Construction.}}
We construct a dynamic, directed, and weighted graph \( G = (V, E, t) \), where \( V \) denotes users, \( E \) represents interactions, and \( t \) is the timestamp associated with each edge. Each post is modeled as a temporal edge whose features are derived from PLM-based text embeddings. If user \( A \) re-posts content from user \( B \), we introduce a directed edge from \( A \) to \( B \), weighted by the frequency of such reposts within a defined time window. Self-loop edges from \( A \) to \( A \) represent original posts, capturing the user's individual content generation. This construction encodes both interpersonal influence and independent ideological expression.
To capture temporal evolution, we partition the graph chronologically into a sequence of snapshots \( G_1, G_2, \ldots, G_{t-1} \), and train models to predict ideological properties in the future snapshot \( G_t \). This forward-looking setup mirrors real-world forecasting scenarios, where a user's historical activity (e.g., during years 1-5) informs predictions about their future ideological stance (e.g., in year 6).

\subsubsection*{\textbf{Temporal Graph Modeling.}}
To capture spatio-temporal dynamics, we evaluate three TGNN models: JODIE, APAN, and TGN. \textbf{JODIE}~\cite{kumar2019predicting} uses coupled RNNs to update user and item embeddings from interactions. A projection operator estimates future user embeddings via time-conditioned scaling:
\begin{equation}
\hat{\mathbf{u}}(t + \Delta t) = (1 + \mathbf{w}_{\Delta t}) \odot \mathbf{u}(t),
\end{equation}
where \( \mathbf{w}_{\Delta t} \) encodes the elapsed-time context.

\textbf{APAN}~\cite{wang2021apan} uses asynchronous updates with an attention-based encoder. Upon interaction, it updates the embedding based on the last state and a ``mailbox'' storing recent neighbor interactions:
\begin{equation}
\mathbf{z}(t) = \text{MLP}(\text{LayerNorm}(\text{MultiHead}(\mathbf{z}(t^-), \mathbf{M}(t)) + \mathbf{z}(t^-))).
\end{equation}

\textbf{TGN}~\cite{tgn_icml_grl2020} maintains time-evolving memory states. Upon interaction between nodes \( i \) and \( j \), it generates a message:
\begin{equation}
\mathbf{m}_i(t) = \text{msg}(\mathbf{s}_i(t^-), \mathbf{s}_j(t^-), \Delta t, \mathbf{e}_{ij}(t)),
\end{equation}
which updates memory via:
$
\mathbf{s}_i(t) = \text{mem}(\mathbf{m}_i(t), \mathbf{s}_i(t^-))
$
.
Final node embeddings \( \mathbf{z}_i(t) \) are computed using temporal attention over neighbors.

\subsubsection*{\textbf{Self-Supervised Link Prediction.}}
As a self-supervised representation-learning stage, temporal link prediction first trains the TGNN to learn node representations by forecasting future edges \( \hat{E}' \) from historical graph sequences. For a candidate node pair \( (u, v) \), the link score is computed as:
\begin{equation}
s(u, v) = \sigma\left(\mathbf{h}_u^\top \mathbf{W} \mathbf{h}_v\right),
\end{equation}
where \( \mathbf{h}_u \) and \( \mathbf{h}_v \) are time-specific embeddings, \( \mathbf{W} \) is a learnable projection matrix, and \( \sigma(\cdot) \) denotes the sigmoid function. This stage produces temporally contextualized node embeddings \( \mathbf{h}_n^t \), which are then used by the downstream ideology-prediction heads.

\subsubsection*{\textbf{Node Classification and Regression.}}
To predict users’ future political ideologies, we attach classification and regression heads to temporal node embeddings \( \mathbf{h}_n^t \) learned via TGNN. PIDN-derived targets are aggregated from each user’s historical posts up to time \( t \). Let \(P_{n,t}\) denote this post set, and let \(\hat{r}_p\) and \(\hat{s}_p\) denote the PIDN-derived relevance label and ideology score for post \(p\), respectively. Political relevance \( R(n)_t \) is defined as the majority post-level relevance label:
\begin{equation}
R(n)_t = \operatorname{mode}\{\hat{r}_p : p \in P_{n,t}\},
\end{equation}
and, for relevant users (\( R(n)_t = 1 \)), let \(P^{+}_{n,t}=\{p\in P_{n,t}:\hat{r}_p=1\}\). Their ideology target is the mean PIDN score over relevant posts:
\begin{equation}
S(n)_t = \frac{1}{|P^{+}_{n,t}|}\sum_{p\in P^{+}_{n,t}}\hat{s}_p.
\end{equation}

To jointly optimize both tasks, we define a unified loss:
\begin{equation}
\mathcal{L}_{\text{PIPN}} = \lambda \mathcal{L}_c + (1 - \lambda)\mathcal{L}_r,
\end{equation}
where \( \lambda \in [0, 1] \) balances classification and regression. This multi-task setup guides the model to learn temporally grounded node representations that capture ideological relevance and intensity.

\vspace{-0.35em}
\section{Experiments}

This section introduces the experiments conducted to validate the framework's efficacy in detecting and predicting political ideologies. We first outline the experimental setup, including datasets, baselines, and implementation details. We then present a detailed evaluation of the \pidn, followed by an analysis of the \pipn. Implementation details are provided in \smref{Appendix~A.2}.

\vspace{-0.35em}
\subsection{Experimental Setup}

\subsubsection{Datasets and Data Preparation}

\paragraph{\textbf{Training Data Construction for \pidn}} \looseness=-1 The \pidn module is trained on a hybrid dataset of politically relevant and irrelevant content. Relevant samples are derived from style-transferred news articles with annotated ideology scores and are labeled according to their political relevance. To balance the data, we include an equal number of daily-life tweets from our Twitter corpus, labeled as politically irrelevant (relevance 0), with a placeholder ideology score that is excluded from the regression objective. To enhance domain adaptation, we add non-political tweets from the \texttt{MajidTweets} dataset~\cite{majid2023tweets}, a large-scale corpus of real-world tweets. These are selected to be disjoint from our Twitter corpus, introducing broader linguistic and stylistic diversity. Importantly, the \pidn regression head is only activated for posts classified as politically relevant, ensuring that ideology scores are not assigned to irrelevant content. We also normalized target scores before training and evaluation.

\paragraph{\textbf{Graph Construction for \pipn}} The experiments for the \pipn module are conducted on two datasets: Truth Social and our large-scale X (Twitter) corpus. The temporal graphs for TGNNs were constructed using posts identified as ``Politically Relevant’’ and subsequently scored by our \pidn-\textbf{\textsc{KL}} model. For the static GNN baselines, we aggregate all temporal interactions into a single, weighted static graph for each dataset. This setup intentionally discards temporal information, allowing us to directly quantify the performance gains attributable to modeling temporal dynamics.

\subsubsection{Models, Baselines, and Evaluation Metrics}

\begin{figure}[t]
    \centering
    \includegraphics[width=0.9\linewidth]{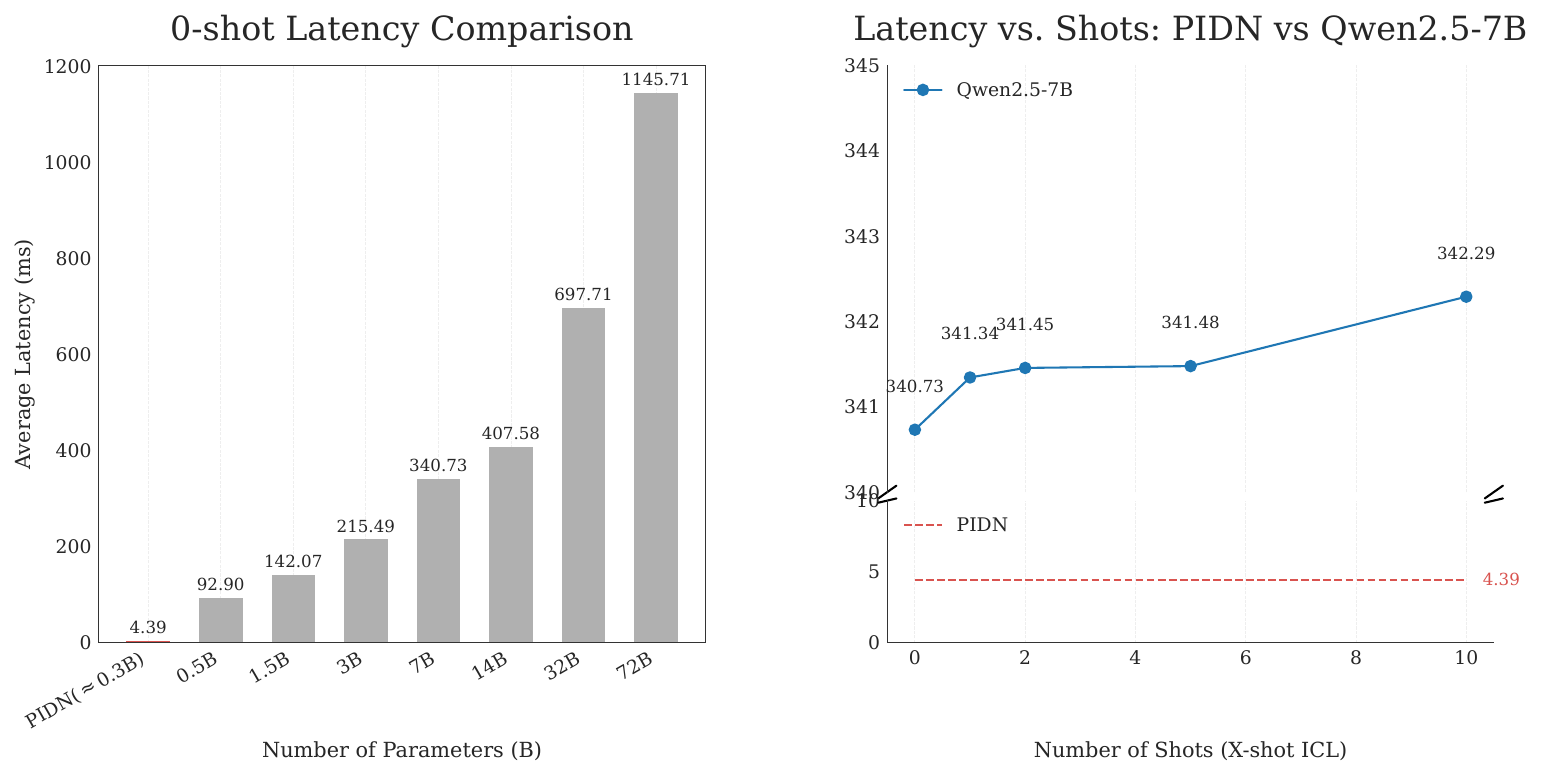}
    \vspace{-0.25em}
    \caption{Inference latency comparison. Left: 0-shot latency of \pidn($\approx$0.3B) vs. LLMs of varying sizes. Right: Latency of \pidn vs. \texttt{Qwen2.5-7B} with increasing ICL shots (0--10). Latencies are reported in milliseconds (ms).}
    \Description{The left bar chart shows PIDN latency at 4.39 milliseconds versus Qwen2.5 models ranging from 92.90 milliseconds for 0.5B to 1145.71 milliseconds for 72B. The right plot shows Qwen2.5-7B latency remaining around 341 to 342 milliseconds across zero to ten shots, while PIDN remains at 4.39 milliseconds.}
    \label{fig:latency_comparison}
\end{figure}

\paragraph{\textbf{Political Ideology Detection (\pidn)}}  
Our evaluation involves: 1) PLM backbone selection, 2) hyperparameter tuning, and 3) ablation of TST and UDA components. We benchmark \pidn against the \texttt{Qwen2.5} model family~\cite{qwen2025qwen25technicalreport}, spanning 0.5B to 72B parameters, in both zero-shot and few-shot ICL settings. Our analysis encompasses the full suite of \texttt{Qwen2.5} models under a zero-shot setting, as well as \texttt{Qwen2.5-7B} with varying numbers of ICL shots, given its widespread use and strong representativeness. All ICL experiments are conducted on 1,000 randomly sampled test instances, repeated five times to ensure robustness.
Evaluation metrics include classification performance—reported as F1 scores for both political/non-political classes and macro-average—as well as class-specific precision and recall for the political class. Regression performance is assessed via Mean Absolute Error (MAE) and Root Mean Squared Error (RMSE), with results summarized in Table~\ref{tab:unified-qwen-analysis}. Inference latency (in milliseconds) is reported in Figure~\ref{fig:latency_comparison}. A comprehensive summary of all performance metrics across models appears in \smref{Tables~6 and 7} of that material.
\looseness=-1 To evaluate the TST module, we adopt two forms of automated assessment. First, we compute semantic similarity between original and style-transferred texts using cosine similarity over sentence embeddings. Second, we apply an \textit{LLM-as-a-Judge} framework to score outputs based on style appropriateness and semantic preservation. All scores are normalized to the $[0,1]$ range for consistency.

\paragraph{\textbf{Political Ideology Prediction (\pipn)}} For the \pipn module, we employ TGNNs as the backbone, which are specifically designed to capture the temporal evolution of interactions within dynamic networks. Given that our datasets model user interactions over time, TGNNs are inherently more suitable for our research objectives. The temporal pipeline is evaluated in two successive stages: 1) self-supervised link prediction, which forecasts future user interactions to learn temporal user embeddings; and 2) downstream political ideology prediction, which combines node classification (determining whether a user's posts are political) and node regression (quantifying the user's ideological score). We conducted a comparative study of three representative TGNNs: TGN~\cite{tgn_icml_grl2020}, JODIE~\cite{kumar2019predicting}, and APAN~\cite{wang2021apan}. We also establish two strong static GNN baselines, GCN~\cite{gcn} and GAT~\cite{velivckovic2017graph}, to validate the necessity of a temporal approach. Performance is measured using average precision (AP) and area under the curve (AUC) for link prediction, and accuracy and MSE for political ideology prediction. The static baselines (GCN, GAT) are trained end-to-end directly on the ideology prediction task, as they do not inherently perform temporal link prediction.

\vspace{-0.35em}
\subsection{Evaluation of the Political Ideology Detection Network (\pidn)}
This section presents an evaluation of \pidn, focusing on its performance and efficiency compared to LLMs, followed by a detailed component analysis and an ablation study.

\begin{table}[t]
  \centering
  \caption{Unified performance analysis of the \texttt{Qwen2.5} model family. \textbf{Top:} cross-model scaling analysis in a zero-shot setting. \textbf{Bottom:} few-shot analysis for the \texttt{Qwen2.5-7B-Instruct} model. Within each block, \best{bold} marks the best and \secbest{underlining} the second best per metric.}
  \label{tab:unified-qwen-analysis}
  \rowcolors{2}{gray!10}{white} %
  \resizebox{\textwidth}{!}{%
    \begin{tabular}{l rrr rr rr}
      \toprule
      & \multicolumn{3}{c}{\textbf{Classification (F1 $\uparrow$)}} & \multicolumn{2}{c}{\textbf{Political Class Detail $\uparrow$}} & \multicolumn{2}{c}{\textbf{Regression Error $\downarrow$}} \\
      \cmidrule(lr){2-4} \cmidrule(lr){5-6} \cmidrule(lr){7-8}
      \textbf{Model / Setting} & Non-political & Political & Macro Avg. & Precision & Recall & MAE & RMSE \\
      \midrule
      \multicolumn{8}{l}{\itshape\textbf{Cross-Model Scaling (0-shot)}}\\
      \midrule
      \texttt{Qwen2.5-0.5B-Instruct} & \pmstd{0.510}{0.022} & \pmstd{0.727}{0.007} & \pmstd{0.618}{0.014} & \pmstd{0.601}{0.007} & \secbest{\pmstd{0.918}{0.006}} & \pmstd{0.413}{0.007} & \pmstd{0.478}{0.005} \\
\texttt{Qwen2.5-3B-Instruct}   & \secbest{\pmstd{0.946}{0.003}} & \secbest{\pmstd{0.943}{0.003}} & \secbest{\pmstd{0.945}{0.003}} & \pmstd{0.945}{0.008} & \best{\pmstd{0.942}{0.006}} & \pmstd{0.283}{0.005} & \pmstd{0.333}{0.005} \\
\texttt{Qwen2.5-7B-Instruct}   & \pmstd{0.927}{0.002} & \pmstd{0.905}{0.003} & \pmstd{0.916}{0.002} & \secbest{\pmstd{0.992}{0.003}} & \pmstd{0.833}{0.004} & \pmstd{0.238}{0.004} & \pmstd{0.288}{0.005} \\
\texttt{Qwen2.5-14B-Instruct}  & \pmstd{0.918}{0.003} & \pmstd{0.900}{0.004} & \pmstd{0.909}{0.003} & \pmstd{0.986}{0.001} & \pmstd{0.827}{0.006} & \pmstd{0.227}{0.004} & \pmstd{0.285}{0.003} \\
\texttt{Qwen2.5-32B-Instruct}  & \secbest{\pmstd{0.946}{0.001}} & \pmstd{0.937}{0.002} & \pmstd{0.942}{0.002} & \pmstd{0.991}{0.001} & \pmstd{0.889}{0.004} & \secbest{\pmstd{0.225}{0.002}} & \secbest{\pmstd{0.283}{0.001}} \\
\texttt{Qwen2.5-72B-Instruct}  & \best{\pmstd{0.955}{0.002}} & \best{\pmstd{0.950}{0.003}} & \best{\pmstd{0.953}{0.002}} & \best{\pmstd{0.994}{0.002}} & \pmstd{0.910}{0.004} & \best{\pmstd{0.216}{0.004}} & \best{\pmstd{0.275}{0.004}} \\
      \midrule
      \multicolumn{8}{l}{\itshape\textbf{Few-Shot for \texttt{Qwen2.5-7B-Instruct}}}\\
      \midrule
      0-shot  & \pmstd{0.927}{0.002} & \pmstd{0.905}{0.003} & \pmstd{0.916}{0.002} & \pmstd{0.992}{0.003} & \pmstd{0.833}{0.004} & \pmstd{0.238}{0.004} & \secbest{\pmstd{0.288}{0.005}} \\
2-shot  & \pmstd{0.936}{0.021} & \pmstd{0.928}{0.026} & \pmstd{0.932}{0.024} & \pmstd{0.986}{0.005} & \pmstd{0.879}{0.049} & \pmstd{0.242}{0.005} & \pmstd{0.292}{0.007} \\
4-shot  & \pmstd{0.920}{0.032} & \pmstd{0.905}{0.044} & \pmstd{0.912}{0.038} & \pmstd{0.986}{0.006} & \pmstd{0.840}{0.079} & \best{\pmstd{0.230}{0.007}} & \best{\pmstd{0.282}{0.012}} \\
6-shot  & \pmstd{0.918}{0.024} & \pmstd{0.903}{0.035} & \pmstd{0.911}{0.030} & \pmstd{0.987}{0.008} & \pmstd{0.835}{0.063} & \pmstd{0.243}{0.007} & \pmstd{0.301}{0.007} \\
8-shot  & \pmstd{0.906}{0.024} & \pmstd{0.886}{0.035} & \pmstd{0.896}{0.029} & \pmstd{0.990}{0.007} & \pmstd{0.804}{0.061} & \pmstd{0.246}{0.007} & \pmstd{0.305}{0.011} \\
10-shot & \pmstd{0.898}{0.019} & \pmstd{0.874}{0.028} & \pmstd{0.886}{0.024} & \pmstd{0.995}{0.002} & \pmstd{0.781}{0.047} & \secbest{\pmstd{0.236}{0.012}} & \pmstd{0.291}{0.015} \\
20-shot & \pmstd{0.923}{0.031} & \pmstd{0.908}{0.043} & \pmstd{0.916}{0.037} & \pmstd{0.995}{0.002} & \pmstd{0.838}{0.074} & \pmstd{0.244}{0.016} & \pmstd{0.306}{0.019} \\
50-shot & \pmstd{0.961}{0.014} & \pmstd{0.958}{0.015} & \pmstd{0.960}{0.014} & \secbest{\pmstd{0.996}{0.003}} & \pmstd{0.924}{0.031} & \pmstd{0.241}{0.018} & \pmstd{0.304}{0.020} \\
100-shot& \secbest{\pmstd{0.964}{0.008}} & \secbest{\pmstd{0.962}{0.009}} & \secbest{\pmstd{0.963}{0.009}} & \best{\pmstd{0.998}{0.000}} & \secbest{\pmstd{0.929}{0.017}} & \pmstd{0.237}{0.010} & \pmstd{0.307}{0.012} \\
200-shot& \best{\pmstd{0.974}{0.008}} & \best{\pmstd{0.973}{0.008}} & \best{\pmstd{0.973}{0.008}} & \best{\pmstd{0.998}{0.002}} & \best{\pmstd{0.950}{0.017}} & \pmstd{0.243}{0.012} & \pmstd{0.315}{0.018} \\
      \bottomrule
    \end{tabular}%
  }%
  \vspace{-0.3em}
\end{table}

\subsubsection{\textbf{Performance and Efficiency vs. Large Language Models}}

\paragraph{\textbf{Performance Analysis.}} To comprehensively evaluate the effectiveness of our model, we first compared our fine-tuned \pidn against the \texttt{Qwen2.5} model family (Table~\ref{tab:unified-qwen-analysis}). For few-shot ICL, each setting was rigorously repeated five times using randomly drawn, class-balanced examples from a held-out pool. Our primary benchmark, \pidn, achieved a near-perfect classification accuracy of 99.25\% and an impressively low regression MSE of 0.0048---an RMSE of 0.069 (Table~\ref{tab:ablation_study}), demonstrating high precision and fine-grained scoring capability.

Cross-model scaling analysis reveals a generally positive, albeit complex and nuanced, trend in zero-shot LLM performance. The largest model, \texttt{Qwen2.5-72B-Instruct}, achieves the highest Macro F1-score of 0.953, establishing it as the top performer. However, the trend is not strictly monotonic, as mid-sized models (7B to 32B) exhibit a strong bias towards precision over recall. They reach near-perfect precision (up to 0.992) but at the cost of lower recall (down to 0.827), reflecting conservative decision boundaries that minimize false positives. For regression, scaling yields more consistent gains; yet, the best LLM's RMSE of 0.275 is still nearly four times worse than \pidn’s 0.069, underscoring the limits of prompting without task-specific tuning.

The few-shot performance of \texttt{Qwen2.5-7B-Instruct} further illustrates these stark limitations. We observe a surprising ``performance trough'': after an initial gain at 2 shots, adding more demonstrations (specifically, 4--10 shots) unexpectedly degrades classification performance, with the F1-score dropping below the zero-shot baseline of 0.916. A consistent improvement over the baseline requires 50 shots or more, which partially corrects the earlier conservative bias by improving recall. Even at its best---a Macro F1 of 0.973 at 200 shots, and a lowest RMSE of 0.282 at 4 shots---ICL still falls well short of our fine-tuned \pidn. This significant gap stems from core methodological differences, as \pidn is explicitly fine-tuned on a large, domain-specific dataset with ideological supervision. This allows it to learn nuanced signals that generic few-shot prompting cannot capture, highlighting the clear superiority of task-specific tuning for precise sociopolitical inference.

\paragraph{\textbf{Latency Analysis.}}
\looseness=-1 To complement the performance evaluation, we conducted a comparative latency analysis to validate the efficiency of our architectural design. We benchmarked our \pidn, built on the \texttt{twhin-bert-base} backbone ($\approx$0.3B parameters), against the same \texttt{Qwen2.5} LLM baselines. Latency is a known limitation of LLMs, especially with larger models or in-context examples. Figure~\ref{fig:latency_comparison} presents the latency comparison. Our \pidn achieves a low inference latency of 4.39 ms. In the left panel, even the smallest \texttt{Qwen2.5} model (0.5B) exhibits a 0-shot latency of 92.90 ms—over 21$\times$ slower than \pidn. Latency increases sharply with model size, reaching 1145.71 ms for the 72B variant, which is more than 260 times slower. The right panel shows the impact of increasing ICL shots on the 7B model. While latency rises only slightly from 340.73 ms (0-shot) to 342.29 ms (10-shot), it remains consistently high—about 78$\times$ slower than \pidn, even under minimal ICL.

\paragraph{\textbf{Discussion on Model Architecture.}}
Our evaluation challenges the common assumption that there is a trade-off between the efficiency of task-specific models and the performance of LLMs. In high-resource settings, our fine-tuned \pidn significantly outperforms LLMs—delivering higher accuracy with orders of magnitude lower inference latency. Even the best ICL setups incur prohibitive computational costs, making LLMs unsuitable for real-time or large-scale deployment.
While \pidn is the clear choice when labeled data is abundant and performance critical, LLMs remain useful in low-resource scenarios where annotation is infeasible. Their zero- and few-shot capabilities offer a practical fallback. Nonetheless, our results strongly support the development of lightweight, specialized models in settings where scalability, precision, and efficiency are most critical. These findings support a broader insight: when conditions allow, architectural specialization often outperforms general-purpose flexibility in real-world analytical tasks.

\subsubsection{Component Analysis and Ablation Studies}

\begin{table}[t]
\centering
\caption{Comparison of PLM backbones and hyperparameter tuning results. Best values are in \textbf{bold}.}
\begin{subtable}[t]{0.59\textwidth}
\centering
\caption{Backbone Comparison}
\rowcolors{2}{gray!10}{white} %
\resizebox{\linewidth}{!}{%
\begin{tabular}{@{}lcc@{}}
\toprule
\textbf{Model} & \textbf{Acc. (Kind) (\%) $\uparrow$} & \textbf{MSE (Score) $\downarrow$} \\
\midrule
\footnotesize\texttt{bert-base-cased}~\cite{devlin2018bert} & \pmstd{95.56}{0.21} & \pmstd{0.0847}{0.0015} \\
\footnotesize\texttt{xlm-roberta-base}~\cite{conneau2019unsupervised} & \pmstd{96.67}{0.15} & \pmstd{0.0686}{0.0011} \\
\footnotesize\texttt{covid-twitter-bert-v2}~\cite{muller2020covid} & \pmstd{95.53}{0.25} & \pmstd{0.0874}{0.0018} \\
\footnotesize\texttt{twhin-bert-base}~\cite{zhang2023twhin} & \textbf{\pmstd{97.79}{0.12}} & \textbf{\pmstd{0.0606}{0.0008}} \\
\bottomrule
\end{tabular}
}
\label{tab:backbone_comparison}
\end{subtable}
\hfill
\begin{subtable}[t]{0.37\textwidth}
\centering
\caption{Hyperparameter Tuning}
\rowcolors{2}{gray!10}{white} %
\resizebox{\linewidth}{!}{%
\begin{tabular}{cccc}
\toprule
\multicolumn{2}{c}{\textbf{Hyperparameters}} & \multicolumn{2}{c}{\textbf{Performance Metrics}} \\
\cmidrule(lr){1-2} \cmidrule(lr){3-4}
\textbf{$\alpha$} & \textbf{$\beta$} & \textbf{Acc. (\%) $\uparrow$} & \textbf{MSE $\downarrow$} \\
\midrule
0.4 & 0.4 & \pmstd{97.12}{0.18} & \pmstd{0.073}{0.002} \\
0.3 & 0.3 & \pmstd{96.98}{0.20} & \pmstd{0.071}{0.002} \\
\textbf{0.4} & \textbf{0.3} & \textbf{\pmstd{97.79}{0.15}} & \textbf{\pmstd{0.067}{0.001}} \\
0.3 & 0.4 & \pmstd{97.52}{0.17} & \pmstd{0.069}{0.002} \\
\bottomrule
\end{tabular}
}
\label{tab:hyperparameter_tuning}
\end{subtable}
\vspace{-1em}
\end{table}

\begin{table}[b]
\centering
\small
\vspace{-0.25em}
\caption{Ablation study of TST and UDA methods. All experiments use the \texttt{twhin-bert-base} backbone and the selected hyperparameter setting ($\alpha=0.4, \beta=0.3$). Best performance is in \textbf{bold}.}
\rowcolors{2}{gray!10}{white} %
\label{tab:ablation_study}
\begin{tabular}{l cc}
\toprule
\textbf{Configuration} & \textbf{Acc. (Kind) (\%) $\uparrow$} & \textbf{MSE (Score) $\downarrow$} \\
\midrule
Backbone Only & \pmstd{97.79}{0.15} & \pmstd{0.0670}{0.0012} \\
\midrule
Backbone + UDA (\textsc{KL}) & \pmstd{98.15}{0.11} & \pmstd{0.0350}{0.0009} \\
Backbone + UDA (\textsc{MMD}) & \pmstd{98.35}{0.10} & \pmstd{0.0320}{0.0010} \\
\midrule
Backbone + TST & \pmstd{99.12}{0.08} & \pmstd{0.0060}{0.0004} \\
\midrule
Backbone + TST + UDA (\textsc{KL}) (\textbf{\pidn-\textsc{KL}}) & \pmstd{99.25}{0.07} & \textbf{\pmstd{0.0048}{0.0003}} \\
Backbone + TST + UDA (\textsc{MMD}) (\textbf{\pidn-\textsc{MMD}}) & \textbf{\pmstd{99.52}{0.05}} & \pmstd{0.0057}{0.0004} \\
\bottomrule
\end{tabular}
\vspace{-1em}
\end{table}

\paragraph{\textbf{Backbone Selection.}} 
We began by evaluating four prominent PLM backbones to identify the most suitable encoder for our task. 
To isolate the backbone’s inherent capacity, this comparison excluded both TST and UDA. 
We used a fixed hyperparameter setting ($\alpha=0.4, \beta=0.2$) for all models, with standard deviations from 5 runs.
As shown in Table~\ref{tab:backbone_comparison}, \texttt{twhin-bert-base} achieved the best results for both political relevance classification (97.79\% accuracy) and ideology score regression (0.0606 MSE). 
This performance underscores the benefit of pre-training on large-scale social media data with user- and network-aware objectives. 
In contrast, although \texttt{covid-twitter-bert-v2} is also social media-oriented, its domain specificity toward COVID-19 limited its effectiveness in capturing the broader political discourse. 
Interestingly, all models yield comparably high classification accuracy, suggesting that identifying whether a tweet expresses political relevance is a relatively easy task for PLMs. 
In other words, classifying the presence of political inclination poses little challenge to modern language models, with minimal variation in performance across different backbones. 
In conclusion, we adopted \texttt{twhin-bert-base} as the backbone for all subsequent experiments.

\paragraph{\textbf{Hyperparameter Tuning.}} Next, we fine-tuned the hyperparameters $\alpha$ and $\beta$, which weight the three terms of the \pidn{} objective $\alpha\mathcal{L}_c+\beta\mathcal{L}_r+(1-\alpha-\beta)\mathcal{L}_{\text{UDA}}$ (classification, regression, and domain alignment; see Figure~\ref{fig:framework}), using our selected \texttt{twhin-bert-base} backbone, again without engaging TST or UDA. Table~\ref{tab:hyperparameter_tuning} reports the four configurations evaluated in this dedicated tuning stage. Among these settings, $\alpha=0.4$ and $\beta=0.3$ achieved the highest classification accuracy (97.79\%) and the lowest MSE (0.067), and we therefore selected this setting for the subsequent ablation study. The $(\alpha=0.4, \beta=0.2)$ setting in Table~\ref{tab:backbone_comparison} was used only to hold the loss weights fixed during the preliminary backbone comparison and was not part of this tuning grid. Accordingly, we do not claim a global optimum across the two stages. The limited variation within the Table~\ref{tab:hyperparameter_tuning} grid suggests that the framework is not highly sensitive to these loss weights.

\looseness=-1 \paragraph{\textbf{Ablation Study.}}
To rigorously evaluate the individual and synergistic contributions of our core components, we conducted a detailed ablation study. Starting with the \texttt{twhin-bert-base} backbone and the selected hyperparameter setting ($\alpha=0.4, \beta=0.3$), we incrementally added TST and the two UDA methods: Kullback-Leibler (\textsc{KL}) divergence and Maximum Mean Discrepancy (\textsc{MMD}). The comprehensive results are presented in Table~\ref{tab:ablation_study}.
The most dramatic improvement stems from TST. By adding TST to the backbone, the regression MSE plummets from 0.0670 to 0.0060—an order-of-magnitude reduction. This starkly demonstrates that bridging the stylistic gap via LLM-based data augmentation is the single most critical factor for achieving high-fidelity ideological scoring.
UDA also yields significant gains. When applied directly to the backbone, MMD reduces the MSE by over 50\% to 0.032, while KL achieves a comparable reduction to 0.035. When combined with TST, both full models deliver the best results among the evaluated configurations. The \pidn-\textbf{\textsc{MMD}} model achieves the highest classification accuracy at 99.52\%, while the \pidn-\textbf{\textsc{KL}} model secures a marginally better regression performance with the lowest MSE of 0.0048. Given that both models demonstrate highly competitive and nearly equivalent performance, the deciding factor shifts to a practical consideration: computational efficiency. As detailed in \smref{Appendix~A.4}, there is a stark contrast in the overhead of the two UDA methods. MMD's quadratic time complexity ($O(N^2 D)$) makes it compute-bound, limiting its scalability with increasing batch size. In contrast, KL divergence's linear complexity ($O(ND)$) makes it significantly more efficient and scalable, posing no practical constraints on large-batch training. Combining top-tier performance with superior computational efficiency, we therefore identify \pidn-\textbf{\textsc{KL}} as our selected configuration.

\begin{figure}[th]
    \centering
    \includegraphics[width=\linewidth]{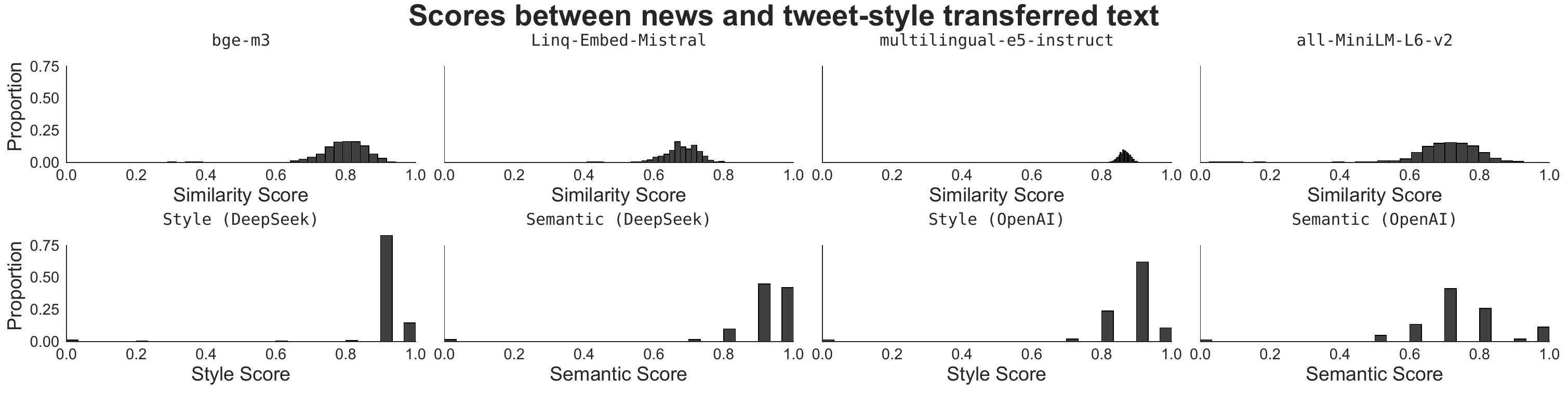}
    \vspace{-0.25em}
    \caption{Evaluation of Text Style Transfer. Top: Histograms of similarity scores between original news and tweet-style outputs, computed using four sentence embedding models. Bottom: Histograms of Style Appropriateness and Semantic Preservation scores (normalized to 0--1) assigned by LLM judges.}
    \Description{Eight histograms evaluate text style transfer. The top row shows news-to-tweet similarity distributions from four embedding models, concentrated mostly between 0.6 and 0.9. The bottom row shows style and semantic scores from DeepSeek and OpenAI judges, with most probability mass at higher scores.}
    \label{fig:tst_evaluation_scores}
\end{figure}

\paragraph{\textbf{Validation of the TST Module.}}
\label{sec:tst_evaluation}
To rigorously assess the contribution and quality of our LLM-based TST module, we conducted a multifaceted evaluation. The goal is to verify that TST effectively adapts formal news into tweet-style social media text while preserving semantic meaning, which is crucial for downstream ideology analysis. The evaluation comprises two complementary components: Semantic Similarity Analysis and \textit{LLM-as-a-Judge} Evaluation.

\begin{itemize}
  \item For Semantic Similarity Analysis, to ensure robustness and reduce model bias, we employed four diverse embedding models: \texttt{bge-m3}~\cite{bge-m3}, \texttt{Linq\allowbreak -Embed\allowbreak -Mistral}~\cite{LinqAIResearch2024}, \texttt{multilingual\allowbreak -e5\allowbreak -large\allowbreak -instruct}~\cite{wang2024multilingual}, and \texttt{all\allowbreak -MiniLM\allowbreak -L6\allowbreak -v2}\footnote{\url{https://huggingface.co/sentence-transformers/all-MiniLM-L6-v2}}. These span multiple architectures (e.g., MiniLM, XLM-R, Mistral, BGE) and sizes (22M--7.1B), providing an architecture-agnostic similarity estimate. High scores across models suggest strong semantic preservation.
  \item For the \textit{LLM-as-a-Judge} Evaluation, we employed \texttt{DeepSeek\allowbreak -V3}~\cite{deepseekai2024deepseekv3technicalreport} and \texttt{GPT-4o}~\cite{hurst2024gpt} as automated evaluators~\cite{zheng2023judging,gu2024survey} to score outputs on \emph{Style Appropriateness} and \emph{Semantic Preservation}, complementing embedding-based metrics.
\end{itemize}

Figure~\ref{fig:tst_evaluation_scores} summarizes the TST evaluation results. The top row shows cosine similarity distributions across four embedding models. \texttt{bge-m3} and \texttt{all-MiniLM-L6-v2} concentrate in the 0.6--0.9 range, indicating solid semantic alignment. \texttt{Linq-Embed-Mistral} yields a similar pattern, slightly skewed lower. Notably, \texttt{multilingual-e5-instruct} peaks near 0.9, suggesting strong semantic consistency. Overall, the results indicate good to excellent semantic preservation.
The bottom row shows LLM-based scores. For \emph{Style Appropriateness}, both \texttt{DeepSeek-V3} and \texttt{GPT-4o} peak around 0.9, confirming alignment with tweet-style norms. For \emph{Semantic Preservation}, \texttt{DeepSeek-V3} yields scores tightly clustered near 0.9--1.0; \texttt{GPT-4o} peaks more broadly around 0.7--0.8, still indicating strong fidelity.
In summary, both embedding- and LLM-based evaluations confirm that our TST module produces tweet-style outputs with preserved meaning. This validates the \pidn pipeline and ensures downstream tasks operate on stylistically consistent, semantically faithful inputs.

\paragraph{\textbf{Bias Analysis and Mitigation}}
Although the TST module is effective, LLM-based style transfer may still bias the data if political cues are shifted unevenly across ideology groups or platforms~\cite{gallegos2023bias}. We focus on three risks: \emph{ideology drift}, where stance-bearing markers are amplified or softened more for one side; \emph{lexical homogenization}, where frequent social-media tokens replace minority or platform-specific expressions and reduce linguistic variety for TGNNs; and \emph{source leakage}, where platform–ideology correlations from pretraining are copied into transferred texts.

We mitigate these effects in three ways. First, we constrain the TST prompt to retain stance phrases and entity mentions, and to modify only surface style (e.g., emojis, contractions, tweet formatting). Second, we perform post-hoc filtering and balancing: pairs with good style but noticeably lower semantic scores from the LLM judges are discarded or regenerated, and \pipn{} batches are balanced across labels and platforms to prevent artefacts from concentrating in one class. This approach aligns with methods for mitigating task-specific biases~\cite{bhardwaj2024catgen}. Third, the evaluation setup itself acts as a safeguard: using four sentence embedding models and two independent LLM-as-a-judge evaluators (\texttt{DeepSeek-V3} and \texttt{GPT-4o}) lowers reliance on any single model, and their agreement indicates that the TST step preserves meaning without systematic distortion.

\vspace{-0.35em}
\subsection{Evaluation of the Political Ideology Prediction Network (\pipn)}
This section evaluates the \pipn module by comparing the evaluated TGNNs with static GNN baselines on link prediction and political ideology prediction.

\paragraph{\textbf{Model Comparison and Analysis}} Table~\ref{table:experiments} presents the results of our comparative analysis. Findings reveal distinct strengths across models, establishing the advantage of temporal modeling.

On the \textbf{Truth Social dataset}, the results reveal a performance trade-off among TGNNs. JODIE achieves the best link-prediction performance among the evaluated models (99.73\% AP and 99.69\% AUC). However, TGN demonstrates significantly better performance on the critical task of ideology classification, achieving the highest accuracy (72.42\%), and it ties with APAN for the lowest regression MSE (0.036). In stark contrast, the static baselines underperform substantially. GAT, the stronger of the two, achieves only 61.81\% accuracy, lagging behind TGN by over 10 percentage points. This considerable performance gap underscores the need for temporal modeling to capture the evolving patterns of ideological expression on this platform.

On the \textbf{X (Twitter) dataset}, TGN consistently outperforms all other models. Here, the advantage of temporal modeling becomes even more pronounced. The static GCN and GAT models reach accuracies of 96.85\% and 95.89\%, respectively—substantially lower than those of the TGNNs (all exceeding 99.9\%). Moreover, the regression performance of the static models is inferior, with MSEs of 0.0793 (GCN) and 0.0629 (GAT), both considerably higher than TGN’s top-performing 0.0560. We attribute the performance gap to the dataset's extensive 16-year span and high interaction density, which allows temporal models to learn evolving ideological trajectories that are inherently lost in the time-aggregated graphs used by static models. This demonstrates that omitting temporal information results in a loss of classification fidelity even with strong topical signals.

\begin{table}[t]
\centering
\small
\vspace{-0.25em}
\caption{Comparative results of static (\xmark) and temporal (\cmark) models. The evaluation covers Link Prediction (AP, AUC) and Ideology Prediction (Accuracy, MSE). Best performance for each metric is in \textbf{bold}; second best is \underline{underlined}. Static models are not applicable (N/A) for the link prediction task.}
\label{table:experiments}
\begin{tabular}{ll c cccc} 
\toprule
\multirow{2}{*}{\textbf{Dataset}} & \multirow{2}{*}{\textbf{Model}} & \multirow{2}{*}{\textbf{Temporal}}
  & \multicolumn{2}{c}{\textbf{Link Prediction}} 
  & \multicolumn{2}{c}{\textbf{Ideology Prediction}} \\
\cmidrule(lr){4-5} \cmidrule(lr){6-7}
 & & & \textbf{AP(\%)\,$\uparrow$} & \textbf{AUC(\%)\,$\uparrow$} & \textbf{Acc.(\%)\,$\uparrow$} & \textbf{MSE\,$\downarrow$} \\
\midrule
\multirow{5}{*}{\textbf{Truth Social}}
  & GCN~\cite{gcn} & \xmark & N/A & N/A & \pmstd{58.35}{0.85} & \pmstd{0.045}{0.002} \\
  & \cellcolor{gray!10}GAT~\cite{velivckovic2017graph} & \cellcolor{gray!10}\xmark & \cellcolor{gray!10}N/A & \cellcolor{gray!10}N/A & \cellcolor{gray!10}\pmstd{61.81}{0.72} & \cellcolor{gray!10}\pmstd{0.043}{0.002} \\
  \cmidrule(lr){2-7}
  & TGN~\cite{tgn_icml_grl2020}   & \cmark & \underline{\pmstd{99.25}{0.08}} & \underline{\pmstd{99.14}{0.09}} & \textbf{\pmstd{72.42}{0.35}} & \textbf{\pmstd{0.036}{0.001}} \\
  & \cellcolor{gray!10}JODIE~\cite{kumar2019predicting} & \cellcolor{gray!10}\cmark & \cellcolor{gray!10}\textbf{\pmstd{99.73}{0.04}} & \cellcolor{gray!10}\textbf{\pmstd{99.69}{0.05}} & \cellcolor{gray!10}\underline{\pmstd{66.28}{0.41}} & \cellcolor{gray!10}\underline{\pmstd{0.039}{0.001}} \\
  & APAN~\cite{wang2021apan}  & \cmark & \pmstd{97.56}{0.15} & \pmstd{97.24}{0.18} & \pmstd{65.14}{0.45} & \textbf{\pmstd{0.036}{0.001}} \\
\midrule
\multirow{5}{*}{\textbf{X (Twitter)}}
  & GCN & \xmark & N/A & N/A & \pmstd{96.85}{0.22} & \pmstd{0.0793}{0.0015} \\
  & \cellcolor{gray!10}GAT & \cellcolor{gray!10}\xmark & \cellcolor{gray!10}N/A & \cellcolor{gray!10}N/A & \cellcolor{gray!10}\pmstd{95.89}{0.28} & \cellcolor{gray!10}\pmstd{0.0629}{0.0011} \\
  \cmidrule(lr){2-7}
  & TGN   & \cmark & \textbf{\pmstd{98.82}{0.05}} & \textbf{\pmstd{98.70}{0.06}} & \textbf{\pmstd{99.99}{0.01}} & \textbf{\pmstd{0.0560}{0.0002}} \\
  & \cellcolor{gray!10}JODIE & \cellcolor{gray!10}\cmark & \cellcolor{gray!10}\pmstd{97.83}{0.09} & \cellcolor{gray!10}\pmstd{97.72}{0.10} & \cellcolor{gray!10}\underline{\pmstd{99.98}{0.01}} & \cellcolor{gray!10}\underline{\pmstd{0.0564}{0.0003}} \\
  & APAN  & \cmark & \underline{\pmstd{98.78}{0.06}} & \underline{\pmstd{98.65}{0.07}} & \underline{\pmstd{99.98}{0.01}} & \underline{\pmstd{0.0564}{0.0004}} \\
\bottomrule
\end{tabular}
\vspace{-1em}
\end{table}

This decision is further solidified by the substantial performance gap exhibited by the static GCN and GAT baselines. Their inability to match the performance of temporal models on our core tasks underscores that a temporal modeling approach is not merely advantageous but essential for accurately predicting political ideology in dynamic networks.

\vspace{-0.35em}
\section{Results and Findings}
This section integrates experimental results to derive novel observations in social media platforms.

\vspace{-0.35em}
\subsection{A Validated Framework for Ideology Detection and Prediction}
\looseness=-1 Our study validates an integrated two-stage framework for analyzing political ideology on social media. In Stage I, the NLP-based \pidn transforms raw textual content into reliable ideological annotations. Building on this foundation, Stage II employs the TGNN-based \pipn to capture the temporal evolution of user interactions and ideological structure dynamics. Together, these components form a robust bridge between micro-level textual expression and macro-level network dynamics.

\vspace{-0.35em}
\subsection{Findings about Echo Chambers and Political Polarization}

We conducted data visualization of social media post content across different levels and objectives. Our goal was to identify shifts in political ideology across social media platforms and explore insights into echo chambers and political polarization.

\subsubsection*{\textbf{Different Social Media Platforms exhibit distinct overarching Political Ideologies.}} 
\begin{figure}[t]
\centering
\begin{subfigure}{0.42\textwidth}
    \centering
    \includegraphics[width=\textwidth]{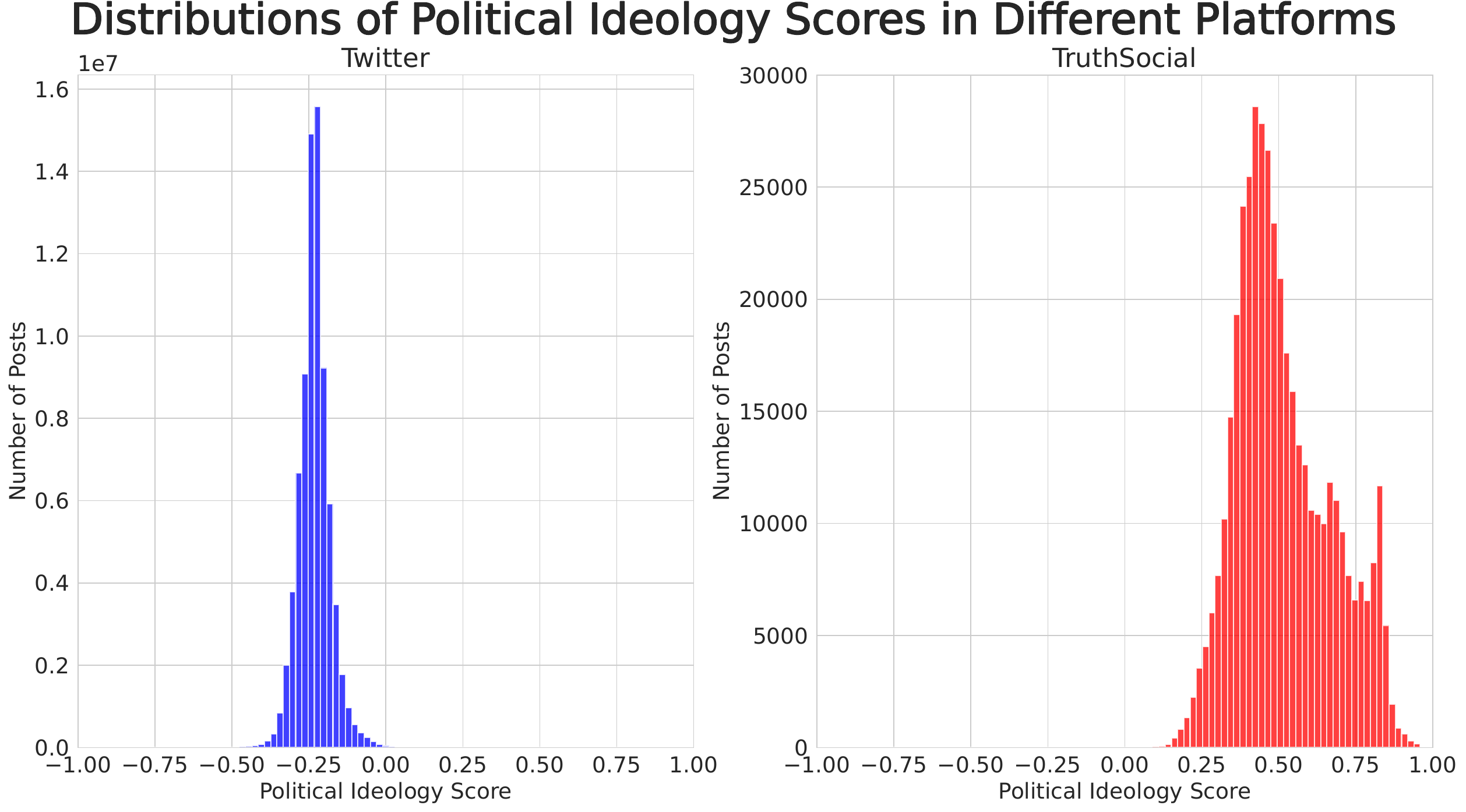}
    \caption{}
    \label{fig:distribution_difference}
\end{subfigure}
\hfill
\begin{subfigure}{0.56\textwidth}
    \centering
    \includegraphics[width=\textwidth]{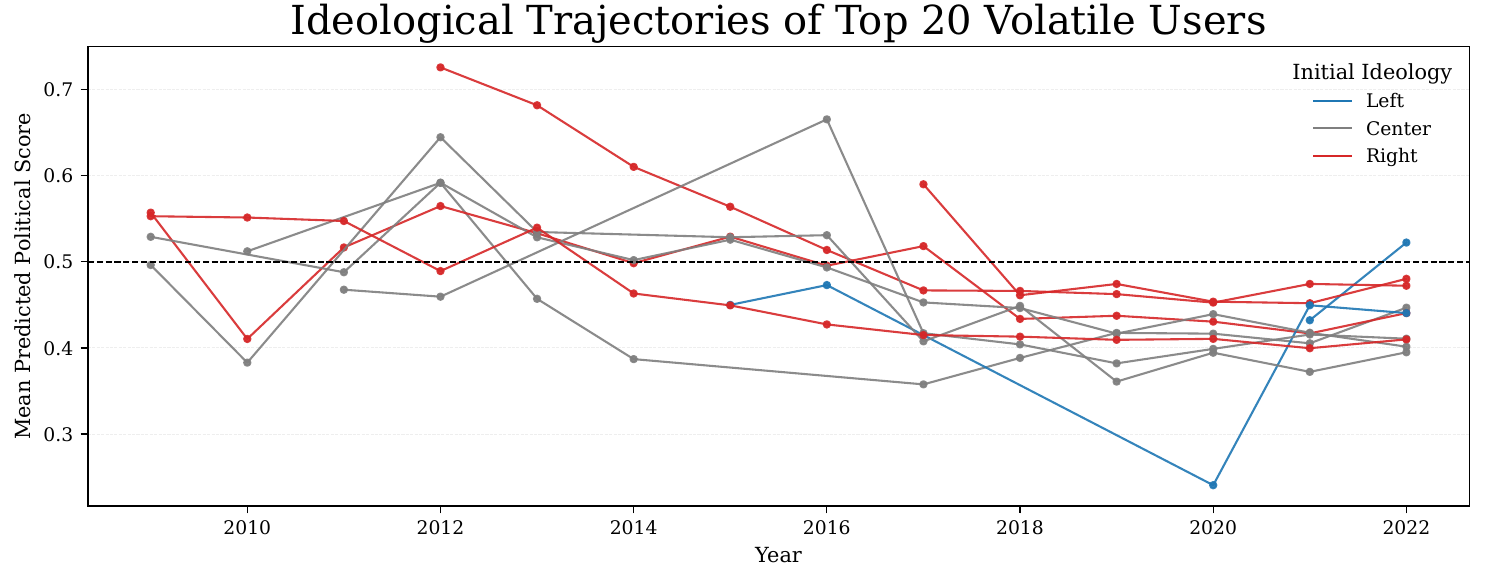}
    \caption{}
    \label{fig:top20_trend}
\end{subfigure}
\vspace{-0.25em}
\caption{Ideological structure and temporal dynamics across platforms. (a) shows how users on the two platforms are distributed along the left–right scale, which reflects cross-platform differences in baseline audience composition. (b) tracks how highly unstable users move on the same scale over time.}
\Description{Panel (a) compares ideology-score histograms: Twitter is concentrated left of zero near -0.25, whereas Truth Social is concentrated on the positive, right-leaning side. Panel (b) plots yearly mean scores for the 20 most volatile Twitter users, colored by initial ideology, with a dashed line at the 0.5 midpoint.}
\label{fig:two_panel_ideology}
\end{figure}

We used \pidn to map each platform's ideological landscape. As illustrated in Figure~\ref{fig:distribution_difference}, despite differences in dataset size, the ideological distributions align with prevailing public perceptions: Twitter exhibits a modest leftward tilt, while Truth Social skews markedly to the right. Twitter, as a widely used mainstream platform, draws from an internet-active American population that tends to lean slightly left. In contrast, Truth Social—founded by former U.S. President Donald J. Trump, a Republican—naturally attracts a predominantly conservative user base.
We further hypothesize that the pronounced rightward shift on Truth Social stems from a combination of factors: the platform’s appeal to influential right-wing figures and its broader user community, which may already possess a mild conservative orientation.

\subsubsection*{\textbf{Ideological trajectories of highly volatile users on Twitter do not support the echo chamber or polarization hypothesis.}}
\looseness=-1 To further examine the dynamics of ideological transformation at the individual level, we identified the top 20 users in the X (Twitter) dataset with the largest spread between their highest and lowest yearly mean ideology scores. To ensure robust estimation, we restricted the analysis to users who posted at least 10 tweets per year and a minimum of 50 tweets in total during the observation period. Each user was assigned an initial political leaning (Left, Center, or Right) based on the mean ideology score in their earliest available year. As shown in Figure~\ref{fig:top20_trend}, we then visualized each user’s ideological trajectory over time, with colors denoting their initial political stance.
We used the Twitter dataset for its broader time span; Truth Social was excluded because its data covers only 2022, too short for the long-term shifts central to this inquiry.

\looseness=-1 Contrary to the classical ``echo chamber'' hypothesis—which posits that users tend to reinforce and radicalize their pre-existing beliefs—we observed complex and multidirectional shifts. Many initially right-leaning users trended towards the ideological center, with several even crossing the midpoint into left-leaning territory over time. This pattern suggests a form of ideological moderation or de-radicalization, rather than an amplification of partisan viewpoints. Meanwhile, the few users in our sample who were initially left-leaning also failed to show a consistent pattern of leftward polarization. One such user, despite significant fluctuations, ultimately returned to an ideological position near their starting point, while another exhibited a clear trajectory toward the right. Most strikingly, users classified as centrist did not remain stable but instead exhibited significant volatility, with their ideological scores fluctuating widely across the political spectrum. In this cohort, an initial centrist position did not predict future stability.

\begin{table*}[t]
  \tiny                                         %
  \centering
  \caption{Case study of three Twitter users illustrating non\hyp{}polarizing trajectories within the U.S.\ political context.  
  For each user, representative tweets are shown from two distinct periods.  
  Key phrases are \textbf{bolded}.}
  \label{tab:case_study}

  \setlength{\tabcolsep}{6pt}       %
  \renewcommand{\arraystretch}{1.25}%

  \begin{tabularx}{\textwidth}{@{} U p{0.4cm} >{\raggedright\arraybackslash}X c @{}}
        \toprule
        \textbf{User} & \textbf{Year} & \textbf{Representative tweet (truncated)} & {\textbf{Score}} \\
        \midrule

        \multirow{5}{*}[-1.75ex]{\centering 1} 
          & \multirow{2}{*}[-1.5ex]{\centering 2009} %
          & Is there such a thing as a \textbf{``Lie Factory''}? If so, its obvious \textbf{right-wing Republicans} \& their talking heads hold a great deal of stock in it. & 0.073 \\[1pt]
          & 
          & \cellcolor{gray!10}3 days left. \textbf{Vote 4 my rescue group 2 save lives}! Retweet! & \cellcolor{gray!10}0.369 \\[1pt]
          \cmidrule(lr){2-4}
          & \multirow{3}{*}[-3.75ex]{\centering 2022} %
          & As you know, currently the \textbf{price per barrel of oil is \$120}. In this chart, note how low the price has to be for corps to profit by drilling. \textbf{It’s the greed}. & 0.515 \\[1pt]
          & 
          & \cellcolor{gray!10}Next time a [Republican] points out that more have died from Covid under [President A] than [President B], remind them that \textbf{correlation doesn’t equal causation}. & \cellcolor{gray!10}0.521 \\[1pt]
          & 
          & This is where I want my taxes to go, \textbf{investing in America and future generations}. A population without access to higher education or drowning in debt cannot compete. & 0.319 \\

        \midrule
        \multirow{6}{*}[-1.75ex]{\centering 2} 
          & \multirow{3}{*}[-0.5ex]{\centering 2017} %
          & There’s been some great addresses about \textbf{\#ScriptureStudy}. What will you begin to study tomorrow? \textbf{\#[ReligiousConference]} & 0.321 \\[1pt]
          &
          & \cellcolor{gray!10}\textbf{\#[ReligiousConference]} \#goals: Study EVERYTHING about Christ \& be a better \textbf{\#Disciple}. Who’s with me? & \cellcolor{gray!10}0.431 \\[1pt]
          &
          & We need our own testimony in these difficult times. Testimonies of others will only carry you so far. \textbf{\#[ChurchLeader]} & 0.240 \\[1pt]
          \cmidrule(lr){2-4}
          & \multirow{3}{*}[-3.0ex]{\centering 2022} %
          & Remember this \textbf{started under [the former President] and is just being continued under [the current President]}. & 0.457 \\[1pt]
          &
          & \cellcolor{gray!10}In 2020 I saw \textbf{Republicans screaming ``you have to vote for [Candidate A]!''} But I also saw \textbf{Dems saying ``you have to vote for [Candidate B] or we will end up with 2016 all over again!''} & \cellcolor{gray!10}0.666 \\[1pt]
          &
          & I've been saying the ``low unemployment'' and ``employee shortage'' have been partly due to the \textbf{great amount of disabled from Covid no longer in the work force}. I was told I was fear-mongering. & 0.455 \\

        \midrule
        \multirow{6}{*}[-1.75ex]{\centering 3} 
          & \multirow{3}{*}[-0.5ex]{\centering 2012} %
          & [The incumbent's] Administration \textbf{Closing 9 Border Patrol Stations} […]. WTF??!!! \textbf{We need MORE, not LESS}! & 0.579 \\[1pt]
          &
          & \cellcolor{gray!10}\textbf{Voter fraud?} Boston reports \textbf{129\% voter turnout}, 79\% for [the incumbent]; 74\% for [a Senate candidate]. & \cellcolor{gray!10}0.518 \\[1pt]
          &
          & In \textbf{\#Iran}, one can be jailed, fined, \& whipped for \textbf{defending \#HumanRights}. Dark reality for [a Christian pastor's] attorney. & 0.363 \\[1pt]
          \cmidrule(lr){2-4}
          & \multirow{3}{*}[-3.0ex]{\centering 2021} %
          & We also have a proud history of our laws \& Constitution being upheld. But today we have an \textbf{illegal two-tiered system} \& a Constitution that the \textbf{Dems \& RINOs don't obey}. & 0.874 \\[1pt]
          &
          & \cellcolor{gray!10}THIS is why people are dying when they go into a hospital for ``C-19 treatment''!!! \textbf{STAY OUT OF HOSPITALS!!!} Signed…a nurse! & \cellcolor{gray!10}0.899 \\[1pt]
          &
          & We're in the \textbf{End Days}, people. Accept Jesus […]. If ever we \textbf{needed Jesus's protection, it's NOW}! & 0.714 \\

        \bottomrule
      \end{tabularx}
  \vspace{-1em}
\end{table*}

\subsubsection*{\textbf{Qualitative analysis reveals diverse, non-polarizing ideological trajectories.}}
To complement our quantitative findings, we conducted a case study to examine user-level ideological dynamics\allowbreak —specifically, whether individuals with high variability in predicted ideology scores exhibit patterns consistent with polarization or show more nuanced shifts. 
We measured each user's ideological volatility as the difference between their maximum and minimum yearly mean scores and selected the top 20 most volatile users. From these, we analyzed three representative cases in detail.
None of these users align with the classical echo chamber hypothesis of unidirectional radicalization; Table~\ref{tab:case_study} presents them.

\begin{itemize}
    \item \looseness=-1 User 1 began using partisan rhetoric in 2009, referring to Republicans as a ``Lie Factory'' (score 0.073) alongside unrelated everyday appeals such as promoting an animal rescue group (score 0.369). By 2022, their tone had shifted toward moderation and analytical reasoning—referencing economic data like ``price per barrel of oil'' (score 0.515) and cautioning that ``correlation doesn't equal causation'' (score 0.521). This trajectory reflects a reduction in rhetorical extremity rather than an intensification of ideology.

    \item \looseness=-1 User 2 transitioned from a non-political participant to a politically engaged yet ideologically balanced commentator. In 2017, their posts focused on religious topics (e.g., \#ScriptureStudy, score 0.321). By 2022, they discussed policies spanning multiple administrations (score 0.457) and observed commonalities in partisan messaging strategies (score 0.666). These expressions reflect a critical, cross-cutting awareness rather than alignment with a single political pole.

    \item \looseness=-1 User 3 maintained consistent conservative views over a decade. In 2012, they raised concerns about border security (score 0.579) and election integrity (score 0.518). By 2021, they continued expressing skepticism toward institutions, citing constitutional violations (score 0.874) and rejecting public health guidance (score 0.899). Though topics shifted, the ideological stance remained stable, reflecting continuity rather than radicalization.

\end{itemize}

Together, these cases challenge the universality of the polarization narrative.

\begin{table*}[t]
\caption{Representative Topics by Political Leaning and Year (2009--2022).}
\label{tab:bertopic_analysis}
\tiny
\resizebox{\textwidth}{!}{%
  \begin{tabularx}{1.1\textwidth}{c >{\columncolor{lightblue}}X >{\columncolor{lightgray}}X >{\columncolor{lightpink}}X}
    \toprule
    \textbf{Year} & \textbf{Left-leaning Topics} & \textbf{Center-leaning Topics} & \textbf{Right-leaning Topics} \\
    \midrule

    2009 &
    \makecell[l]{\texttt{[hcr, health, reform]} \\ \texttt{[copenhagen, cop15]}} &
    \makecell[l]{\texttt{[stimulus, job, economy]} \\ \texttt{[hcr, health, reform]} \\ \texttt{[iran, iranelection]}} &
    \makecell[l]{\texttt{[spending, stimulus, govt]} \\ \texttt{[hcr, gop, tcot]} \\ \texttt{[iran, iranelection]}} \\
    \midrule

    2010 &
    \makecell[l]{\texttt{[health, reform, hcr]} \\ \texttt{[haiti, earthquake]}} &
    \makecell[l]{\texttt{[oil, bp, oilspill]} \\ \texttt{[haiti]} \\ \texttt{[election, mn2010, tea, teaparty]}} &
    \makecell[l]{\texttt{[hcr, obamacare, repeal]} \\ \texttt{[tea, party, teaparty]} \\ \texttt{[flgov, scott, spending]}} \\
    \midrule

    2011 &
    \makecell[l]{\texttt{[egypt, libya]} \\ \texttt{[medicare, cuts, deficit]} \\ \texttt{[wiunion]}} &
    \makecell[l]{\texttt{[debt, debtceiling, boehner]} \\ \texttt{[egypt, mubarak, libya]} \\ \texttt{[bin laden, pakistan]}} &
    \makecell[l]{\texttt{[libya, nato, war]} \\ \texttt{[wiunion, gov, walker]} \\ \texttt{[debt, obama, gop]}} \\
    \midrule

    2012 &
    \makecell[l]{\texttt{[waronwomen, women]} \\ \texttt{[romney, mitt, returns]} \\ \texttt{[health, medicare]}} &
    \makecell[l]{\texttt{[obama, romney, debate]} \\ \texttt{[syria]} \\ \texttt{[gun, guncontrol, newtown, nra]}} &
    \makecell[l]{\texttt{[libya, benghazi, attack]} \\ \texttt{[obama, romney, debate]} \\ \texttt{[gun, shooting, aurora]}} \\
    \midrule

    2013 &
    \makecell[l]{\texttt{[standwithwendy, txlege, abortion]} \\ \texttt{[gun, guns, guncontrol]} \\ \texttt{[blacklivesmatter]}} &
    \makecell[l]{\texttt{[snowden, leak, nsa]} \\ \texttt{[boston, marathon, bombing]} \\ \texttt{[shutdown, government]}} &
    \makecell[l]{\texttt{[irs, lerner, targeting]} \\ \texttt{[benghazi, cia]} \\ \texttt{[syria, assad, military]}} \\
    \midrule

    2014 &
    \makecell[l]{\texttt{[ferguson, police, racism]} \\ \texttt{[acaworks, aca]} \\ \texttt{[climate, change]}} &
    \makecell[l]{\texttt{[ferguson, michael brown]} \\ \texttt{[ukraine, russia, putin]} \\ \texttt{[ebola]} \\ \texttt{[mh370]}} &
    \makecell[l]{\texttt{[isis, syria, iraq]} \\ \texttt{[benghazi]} \\ \texttt{[bridgegate, christie]} \\ \texttt{[irs, lerner, emails]}} \\
    \midrule

    2015 &
    \makecell[l]{\texttt{[blacklivesmatter]} \\ \texttt{[demdebate, berniesanders]} \\ \texttt{[scottwalker, antiunion]}} &
    \makecell[l]{\texttt{[trump, realdonaldtrump, gopdebate]} \\ \texttt{[iran, irandeal]} \\ \texttt{[paris, parisattacks]}} &
    \makecell[l]{\texttt{[iran, deal, irandeal]} \\ \texttt{[plannedparenthood, standwithpp]} \\ \texttt{[gopdebate, trump]}} \\
    \midrule

    2016 &
    \makecell[l]{\texttt{[feelthebern, berniesanders]} \\ \texttt{[flintwatercrisis]} \\ \texttt{[demsinphilly]}} &
    \makecell[l]{\texttt{[trump, hillaryclinton, debate]} \\ \texttt{[brexit, euref]} \\ \texttt{[scalia, scotus]}} &
    \makecell[l]{\texttt{[trump, realdonaldtrump, maga]} \\ \texttt{[wikileaks, emails, clinton]} \\ \texttt{[benghazi]}} \\
    \midrule

    2017 &
    \makecell[l]{\texttt{[muslimban, travelban, nobannowall]} \\ \texttt{[aca, healthcare]} \\ \texttt{[impeachment]}} &
    \makecell[l]{\texttt{[comey, mueller, trumprussia]} \\ \texttt{[charlottesville]} \\ \texttt{[metoo, sexual harassment]}} &
    \makecell[l]{\texttt{[gorsuch, scotus]} \\ \texttt{[fakenews, cnn]} \\ \texttt{[tax, taxreform]} \\ \texttt{[obamacare, repeal]}} \\
    \midrule

    2018 &
    \makecell[l]{\texttt{[guncontrolnow, nra, parkland]} \\ \texttt{[bluewave, midterms]} \\ \texttt{[metoo, kavanaugh]}} &
    \makecell[l]{\texttt{[kavanaugh, ford, kavanaughhearings]} \\ \texttt{[border, children, familiesbelongtogether]} \\ \texttt{[khashoggi]}} &
    \makecell[l]{\texttt{[witch hunt, mueller, fbi]} \\ \texttt{[caravan, border, buildthewall]} \\ \texttt{[kavanaugh, confirmkavanaugh]}} \\
    \midrule

    2019 &
    \makecell[l]{\texttt{[impeachment, impeach, repadamschiff]} \\ \texttt{[greennewdeal]} \\ \texttt{[demdebate, warren]}} &
    \makecell[l]{\texttt{[impeachment, ukraine, whistleblower]} \\ \texttt{[muellerreport, obstruction]} \\ \texttt{[hong kong]}} &
    \makecell[l]{\texttt{[impeachment, sham, hoax]} \\ \texttt{[biden, hunter, ukraine]} \\ \texttt{[border, wall, emergency]}} \\
    \midrule

    2020 &
    \makecell[l]{\texttt{[covid19, pandemic, trumpvirus]} \\ \texttt{[george floyd, blm, defundthepolice]} \\ \texttt{[usps, mailin]}} &
    \makecell[l]{\texttt{[coronavirus, covid19]} \\ \texttt{[biden, trump, election]} \\ \texttt{[george floyd, protests]}} &
    \makecell[l]{\texttt{[obamagate, spygate]} \\ \texttt{[reopenamerica, reopen]} \\ \texttt{[voterfraud, electionfraud, stopthesteal]}} \\
    \midrule

    2021 &
    \makecell[l]{\texttt{[jan 6, insurrection, coup]} \\ \texttt{[votingrights, filibuster]} \\ \texttt{[buildbackbetter, bbb]}} &
    \makecell[l]{\texttt{[capitol, riot, jan 6]} \\ \texttt{[afghanistan, withdrawal]} \\ \texttt{[vaccine, mandate, delta]}} &
    \makecell[l]{\texttt{[crt, critical race theory]} \\ \texttt{[arizona, audit]} \\ \texttt{[antimask, antivax, bordercrisis]}} \\
    \midrule

    2022 &
    \makecell[l]{\texttt{[abortion, codifyroe, womensrights]} \\ \texttt{[uvalde, gunreform]} \\ \texttt{[ukraine, standwithukraine]}} &
    \makecell[l]{\texttt{[ukraine, war, russia]} \\ \texttt{[maralago, search, documents]} \\ \texttt{[roe, wade, overturn]}} &
    \makecell[l]{\texttt{[inflation, gas]} \\ \texttt{[fbi, doj]} \\ \texttt{[border, migrants]} \\ \texttt{[hunter, laptop]}} \\
    \bottomrule
  \end{tabularx}%
} %
\vspace{-1em}
\end{table*}

\subsubsection*{\textbf{Comparative Thematic Analysis Reveals Divergent Discursive Arenas}}

\looseness=-1 To complement our quantitative findings, we conducted a thematic analysis of political discourse on Twitter and Truth Social—two ideologically distinct platforms. We grouped users into Left-leaning, Centrist, and Right-leaning categories based on aggregated post-level ideology scores, and applied BERTopic~\cite{grootendorst2022bertopic} to extract dominant themes. BERTopic embeds documents contextually, reduces their dimensionality, clusters them by density, and extracts representative keywords per topic. Results are shown in Table~\ref{tab:bertopic_analysis} (Twitter, 2009--2022) and Table~\ref{tab:bertopic_analysis_truthsocial} (Truth Social, 2022).

\looseness=-1 Twitter reveals a dynamic we term a ``shared agenda with competing frames.'' Across the political spectrum, users engage with the same high-salience topics—from \texttt{[hcr]} in 2009 to \texttt{[ukraine, war]} in 2022—suggesting a common baseline of topical attention. However, ideological divergence emerges in framing. For example, during the 2019 impeachment, the Left referenced institutional actors (\texttt{[repadamschiff]}), while the Right used delegitimizing terms (\texttt{[sham, hoax]}). Similarly, the events of January 6th surfaced as \texttt{[insurrection, coup]} on the Left and \texttt{[capitol, riot]} at the Center, while the Right's 2021 agenda instead foregrounded \texttt{[arizona, audit]} and \texttt{[crt, critical race theory]}. This is a contested discursive space, not a set of isolated echo chambers.

\looseness=-1 In contrast, Truth Social exhibits a ``high-conflict ideological arena'' with a right-shifted discursive center. Themes categorized as far-right on Twitter—e.g., \texttt{[hunter, laptop]}—appear under the ``Centrist'' label on Truth Social, alongside claims such as \texttt{[2000mules, ballot]} that have no Twitter counterpart, reflecting that partisan narratives constitute the platform’s mainstream. Rhetorical intensity is notably elevated: economic concerns are framed as \texttt{[bidenflation]}, the Mar-a-Lago search is portrayed as a partisan \texttt{[witchhunt]} by a \texttt{[corrupt]} FBI, and the term \texttt{[crimefamily]} denounces the Bidens. Such vocabulary suggests a shift from political discourse to moral indictment.
\looseness=-1 While Left-leaning users exist on Truth Social, they act more as ideological foils than dialogue participants. Topics like \texttt{[jan6, hearings]} and \texttt{[gun control]} appear, but often in oppositional framing. Instead of reducing polarization, their presence intensifies it by provoking adversarial responses. This dynamic is compounded by conspiratorial clusters—e.g., \texttt{[stoptheshots, pureblood]}—which anchor its ideological extremity.

\begin{table*}[t]
\caption{Representative Topics on Truth Social by Political Leaning (2022).}
\label{tab:bertopic_analysis_truthsocial}
\tiny
\resizebox{\textwidth}{!}{%
  \begin{tabularx}{1.1\textwidth}{c >{\columncolor{lightblue}}X >{\columncolor{lightgray}}X >{\columncolor{lightpink}}X}
    \toprule
    \textbf{Year} & \textbf{Left-leaning Topics} & \textbf{Center-leaning Topics} & \textbf{Right-leaning Topics} \\
    \midrule

    2022 &
    \makecell[l]{%
        \texttt{[abortion, roe, wade, overturned]} \\ 
        \texttt{[uvalde, shooting, gun, control]} \\ 
        \texttt{[ukraine, standwithukraine, kyiv]} \\
        \texttt{[jan6, committee, hearings, insurrection]} \\
        \texttt{[maralago, raid, fbi]}
    } &
    \makecell[l]{%
        \texttt{[ukraine, russia, war, putin]} \\ 
        \texttt{[maralago, raid, fbi, doj, warrant]} \\ 
        \texttt{[roe, wade, overturned, supreme, court]} \\
        \texttt{[inflation, recession, economy, prices]} \\
        \texttt{[hunter, laptop, bidens, fbi]} \\
        \texttt{[2000mules, ballot, harvesting, fraud]}
    } &
    \makecell[l]{%
        \texttt{[inflation, gas, prices, bidenflation]} \\
        \texttt{[fbi, doj, raid, corruption, witchhunt]} \\
        \texttt{[border, crisis, migrants, illegal]} \\
        \texttt{[hunter, laptop, biden, crimefamily]} \\
        \texttt{[stoptheshots, diedsuddenly, pureblood]} \\
        \texttt{[abortion, prolife, unborn, baby]}
    } \\
    \bottomrule
  \end{tabularx}%
} %
\vspace{-1em}
\end{table*}

\looseness=-1 Overall, polarization manifests differently across platforms: Twitter hosts a fragmented but interconnected public sphere where competing frames coexist, whereas Truth Social functions as a self-reinforcing enclave whose right-shifted center, moralized rhetoric, and reactive dynamics amplify division rather than bridge it.

\vspace{-0.35em}
\section{Conclusion}
\looseness=-1 This paper presents \framework, a unified framework that integrates natural language processing with temporal graph neural networks for tracking political ideologies on social media. Methodologically, we provide a robust, end-to-end pipeline that effectively bridges the domain gap between news and social content, modeling the dynamic evolution of user interactions. We also contribute two large-scale, publicly released datasets from X (formerly Twitter) and allsides.com. Empirically, our analysis of these platforms reveals that the most ideologically volatile users tend to converge toward the center, a finding contrary to widely held assumptions about increasing online polarization. Future work includes extending to multi-dimensional ideological models and cross-platform comparative studies.

\begin{acks}
This work was supported in part by the National Natural Science Foundation of China (Grant Nos. 92370204 and 62506348), the National Key R\&D Program of China (Grant No. 2023YFF0725001), the New Generation Artificial Intelligence-National Science and Technology Major Project (Grant No. 2025ZD0122601), Guangdong Provincial Key Laboratory of Frontier Basic Science for All-domain Intelligence, the Guangdong Basic and Applied Basic Research Foundation (Grant No. 2023B1515120057), the Key-Area Special Project of Guangdong Provincial Ordinary Universities (Grant No. 2024ZDZX1007), the Natural Science Foundation of Anhui Province (Grant No. 2508085QF211), and the Opening Foundation of State Key Laboratory of Cognitive Intelligence, iFLYTEK (Grant No. COGOS-2025HE02).
\end{acks}
\clearpage

\input{supplementary}

\clearpage

\bibliographystyle{ACM-Reference-Format}
\bibliography{sample-base}

\end{document}

%% file: supplementary.tex
\clearpage
\begin{center}
{\LARGE\bfseries Supplementary Material}
\end{center}

\setcounter{table}{0}
\setcounter{figure}{0}
\setcounter{equation}{0}

\noindent This document is the online-only supplementary material for the article
``Against Political Polarization: A Unified Framework for Tracing Evolving Political
Ideologies on Social Media.'' It collects dataset details, implementation settings,
prompt templates, complexity analysis, and the full per-model results referenced from
the main text. Section, table, and figure numbers below are those cited in the main
text as belonging to the supplementary material.

\appendix
\section{Appendix}
\etocsetnexttocdepth{4}

\localtableofcontents
\subsection{Details of the Datasets}
\subsubsection{Data Privacy}
\label{app:data_privacy}
Given the sensitivity of political ideology, privacy is central to our data handling. All news data were sourced from public websites. Social media data were anonymized to meet ethical standards. On X and Truth Social, personal identifiers (e.g., names, handles) were replaced with numeric IDs. Only public posts were retained; no private user data was used.

\noindent \textbf{Data Release Format.} To respect the terms of the underlying sources, the released datasets are distributed in a \emph{dehydrated} form. For the X (Twitter) and Truth Social corpora we release post and user identifiers together with our derived annotations (political relevance labels, ideology scores, and the temporal interaction edges used to build the graphs), rather than the original post text; users can rehydrate the text through the respective platform APIs subject to their terms of service. For the news corpus we release article URLs, outlet names, publication dates, and the AllSides ideology score of each outlet, rather than the full article text, which remains the property of the publishing outlets. The AllSides ratings redistributed in this way are covered by the CC~BY-NC~4.0 license and are provided with the attribution required by that license; the released collection is therefore available for noncommercial research use only. Scripts for rehydration and for reproducing every table in the paper are provided in the repository at \url{https://github.com/yeahjack/TSN4PI}.
\subsubsection{The allsides.com Dataset}
The overview of the allsides.com dataset is as follows:
\begin{table}[h]
\centering
\small
\caption{Overview of the allsides.com news with annotated political ideology dataset.}
\vspace{-1em}
\rowcolors{2}{gray!10}{white} %
\begin{tabular}{@{}lrr@{}}
\toprule
\textbf{Characteristic} & \multicolumn{1}{l}{\textbf{\# of News Articles}} & \multicolumn{1}{l}{\textbf{\# of Media Outlets}} \\ 
\midrule
Total Articles/Outlets & 233,570 & 466 \\
\midrule
Political Ideology & & \\
\midrule
\quad Center & 76,023 & 158 \\

\quad Lean Left & 53,532 & 106 \\

\quad Right & 45,506 & 75 \\

\quad Lean Right & 32,555 & 62 \\

\quad Left & 25,954 & 65 \\
\bottomrule
\end{tabular}
\label{tab:allsides_overview}
\vspace{-1em}
\end{table}

\noindent \textbf{Data Preprocessing.}
\label{app:allsides_preprocessing}
As with the X dataset described below, we retained only English articles to ensure linguistic consistency. We also normalized ideology scores to align with other framework components. For an article \(N\) with original score \(S_{\text{raw}} \in [-6,6]\), the adjusted score \(S_{\text{adjusted}}\) is computed as: 
\begin{equation}S_{\text{adjusted}}(N) = \frac{S_{\text{raw}} + 6}{12}.\end{equation}

This normalization maps the score to the $[0,1]$ interval, facilitating uniform scale interpretation during model training and evaluation.

\begin{figure}[htbp]
    \centering
    \includegraphics[width=1.0\linewidth]{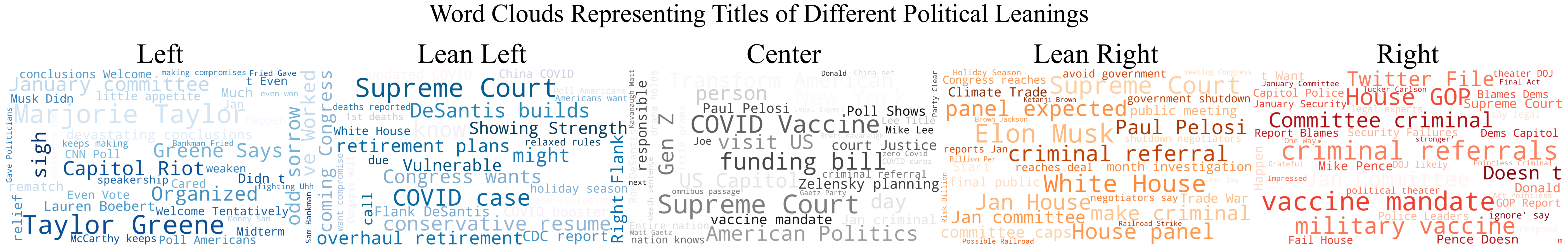}
    \caption{Word Clouds Highlighting Common Terms in News Reports Titles by Political Leanings. The visualization spans from `Left' to `Right', with `Center' in the middle, reflecting the language usage patterns in political news reporting.}
    \Description{Five side-by-side word clouds compare title terms for Left, Lean Left, Center, Lean Right, and Right outlets. Prominent phrases include Capitol Riot and Taylor Greene on the left, COVID vaccine and Supreme Court in the center, and criminal referrals and vaccine mandate on the right.}
    \label{fig:allsides_titles_wordcloud}
\end{figure}

We also created word clouds for news titles from different political ideology factions, as shown in Figure~\ref{fig:allsides_titles_wordcloud}. We found that media with a left-leaning bias generally focus on topics such as the Capitol riot, the Supreme Court, and COVID, while right-leaning media often focus on Elon Musk, Paul Pelosi, the Committee Criminal, and vaccine mandates, with center media reporting on both. This reflects not only the different focus points of media across the political spectrum but also the distinct use of language in reporting the same topics, aligning with our research hypothesis: even for the same events, the choice of words can still reveal their political ideologies.

\subsubsection{The X (Twitter) Dataset}
\label{app:x_dataset}
\noindent \textbf{Data Preprocessing.} To ensure linguistic consistency with our analytical focus of presidential elections in the United States, we retained only English-language tweets. Language filtering was performed using the \texttt{fasttext-langdetect} library~\cite{joulin2016fasttext, joulin2016bag}.

\subsubsection{The Truth Social Dataset}
\label{app:truth_social_dataset}
\noindent \textbf{Data Preprocessing.} Preliminary cleaning and preprocessing were applied to the raw dataset. Given that some ``Truths'' had missing or misaligned timestamps, we utilized web scraping tools to supplement and correct the timestamps for these accessible ``Truths.''

\subsection{Implementation Details}
\label{app:implementation_details}
Models were trained with a 90/10 train-test split. For \pidn, we trained for 3~epochs with a learning rate of $2\times10^{-5}$ and a weight decay of 0.01. For the TST task, we used the \texttt{Llama-3.1-8B-Instruct} model to generate style-transferred texts with a temperature of~0.7, \texttt{top\_p} of~0.95, and a maximum generation length of 512 tokens, terminating on the ``\texttt{</Tweet>}'' token. 
The prompts for directly using the LLM-based method to assess political relevance and assigning scores, TST and \textit{LLM-as-a-judge} for evaluation are provided in Tables~\ref{prompt:llm_based_method}, \ref{prompt:tst}, and~\ref{prompt:llm_judge_tst}, respectively. 
To facilitate automatic evaluation, we leverage tool calling to parse model responses into a structured JSON format. 
For \pidn, we used $(\alpha=0.4, \beta=0.2)$ during the preliminary backbone comparison. After the dedicated tuning stage reported in the main paper, we selected $(\alpha=0.4, \beta=0.3)$ for the ablation study and final models. For \pipn, we adopted the default sampling, memory, and network configurations provided by TGL~\cite{zhou2022tgl} for the different TGNNs. Mixed-precision training and inference were adopted for efficiency. All experiments were conducted on eight machines, each equipped with 8 \textit{Nvidia A800} GPUs (80GB VRAM each), running \textit{Ubuntu 22.04} with \textit{Python 3.11}. We used the \texttt{vllm}~\cite{kwon2023efficient} framework for text generation tasks, and \texttt{TGL}~\cite{zhou2022tgl} for TGNNs. For static graph tasks, we used \texttt{PyTorch Geometric}~\cite{fey2019fast}. The training and inference were performed using \texttt{PyTorch}~2.4.0.

\subsection{Prompt Details}
\subsubsection*{\textbf{Prompt for LLM-based method}} Here we provide the prompt template used for the LLM-based method, which is designed to generate relevance and scores for posts. We enabled tool calling to allow the model to output structured results after reasoning about the political relevance and leaning of a given tweet.

\begin{center}
\refstepcounter{table}
\begin{tcolorbox}[
    breakable,
    standard jigsaw,
    title=Prompt Template for Political Tweet Analysis with Function Calling,
    width=1.0\linewidth,
    halign=left,
    opacityback=0,
    label=prompt:llm_based_method,
    fontupper=\footnotesize,
    colback=white,
    boxrule=0.5pt,
    arc=2mm,
    top=2mm, bottom=2mm,
    toptitle=1mm, bottomtitle=1mm,
]

\textbf{System Prompt:}

You are an expert political analyst specializing in U.S. politics. Your task is to analyze a given tweet and provide a structured analysis by using the \texttt{record\_political\_analysis} tool. Carefully determine the following parameters for the tool:
\begin{itemize}
    \item \texttt{relevance}: Set to '1' if the tweet is politically relevant to the U.S. context, '0' otherwise.
    \item \texttt{score}: If relevant, assign a float from 0.0 (left-leaning) to 1.0 (right-leaning). This parameter should only be provided for relevant tweets.
    \item \texttt{reasoning}: Provide a clear, step-by-step explanation for your decisions. Escape all double quotes with \textbackslash{}.
\end{itemize}
You must directly use the \texttt{record\_political\_analysis} function to output your findings.

\noindent\rule{\linewidth}{0.4pt}
\textbf{Guidelines \& Tool Schema (\texttt{record\_political\_analysis}):}
\begin{Verbatim}[commandchars=\\\{\}]
\textbf{Function Name}: record_political_analysis
\textbf{Description}: Records the political analysis of a tweet after careful reasoning.

\textbf{Parameters}:
- `reasoning` (string, required): Detailed step-by-step thinking.
- `relevance` (string, required): '1' if political, '0' otherwise.
- `score` (number, optional): Score 0.0-1.0. Only required if relevance=='1'.
\end{Verbatim}

\noindent\rule{\linewidth}{0.4pt}
\textbf{Example Input Tweet}: \textit{BREAKING: The Senate just passed the new infrastructure bill. A major win for the administration, but critics on both sides are already voicing concerns about the spending.}

\textbf{Example Output (Tool Call Arguments)}:
\begin{lstlisting}[style=jsonstyle]
{
  "reasoning": "The tweet discusses a major piece of U.S. legislation, the infrastructure bill, and mentions the Senate and the administration. This is clearly a political topic. The tweet presents a balanced view, noting it as a 'win' but also mentioning 'critics on both sides', suggesting a centrist take. Therefore, the score is set to 0.5.",
  "relevance": "1",
  "score": 0.5
}
\end{lstlisting}

\noindent\rule{\linewidth}{0.4pt}
\textbf{Input Format:} \texttt{Analyze this tweet:\string\n\string\n\{tweet\_text\}}

\textbf{Start of Response:} The model should directly output the tool call. The arguments should be a single, well-formed JSON string.
\end{tcolorbox}
\end{center}

\subsubsection*{\textbf{Prompt for Text Style Transfer.}} Here we provide the prompt template used for the text style transfer (TST) task, which is designed to convert news articles into concise tweets. This prompt is crucial for training our model to understand the nuances of tweet generation from longer news content.

\vspace{-0.5em}
\begin{center}
\refstepcounter{table}
\begin{tcolorbox}[
    breakable,
    standard jigsaw,
    title=Prompt Template for News-to-Tweet Style Transfer,
    width=1.0\linewidth,
    halign=left,
    opacityback=0,
    label=prompt:tst,
    fontupper=\footnotesize,
    colback=white,
    boxrule=0.5pt,
    arc=2mm,
    top=2mm, bottom=2mm,
    toptitle=1mm, bottomtitle=1mm,
]

\textbf{System Prompt:}

You will be given a news report. Your task is to rewrite it as a tweet that captures the core message.

\noindent\rule{\linewidth}{0.4pt}
\textbf{Guidelines:}
\begin{Verbatim}[commandchars=\\\{\}]
  Be \textbf{concise} and stay within 280 characters.
  Use common \textbf{abbreviations} (e.g., \textit{ICYMI}, \textit{IMO}).
  Add \textbf{hashtags} to highlight key topics (e.g., \#Breaking).
  Mention relevant accounts using \textbf{@} when applicable.
  Always produce a non-empty tweet in \textbf{ENGLISH}.
  End your generated tweet with the token: \texttt{</Tweet>}
\end{Verbatim}

\noindent\rule{\linewidth}{0.4pt}
\textbf{Example News}: \textit{In an extraordinary face-to-face meeting at the White House on Wednesday, President Joe Biden told Ukrainian President Volodymyr Zelenskyy that\ldots}

\textbf{Example Tweet}: \textit{President Biden meets with Ukrainian President Zelenskyy at the White House, pledging ``unequivocal and unbending support'' to Ukraine. \#UkraineWar \#WhiteHouseMeeting}

\noindent\rule{\linewidth}{0.4pt}
\textbf{Input Format:} \texttt{<Tweet>\{news\_text\}</Tweet>}

\textbf{Start of Response:} \texttt{<Tweet>}
\end{tcolorbox}
\end{center}

\subsubsection*{\textbf{Prompt for Text Style Transfer Evaluation.}} This prompt is used to evaluate the quality of the generated tweets from the TST task by \textit{LLM-as-a-judge}. It is designed to assess the relevance and quality of the generated tweet in relation to the original news article. We enabled tool calling to allow the model to output structured scores after reasoning about the tweet's style and semantic preservation.

\begin{center}
\refstepcounter{table}
\begin{tcolorbox}[
    breakable,
    standard jigsaw,
    title=Prompt for \textit{LLM-as-a-Judge} Evaluation of Text Style Transfer,
    width=1.0\linewidth,
    halign=left,
    opacityback=0,
    label=prompt:llm_judge_tst,
    fontupper=\footnotesize,
    colback=white,
    boxrule=0.5pt,
    arc=2mm,
    top=2mm, bottom=2mm,
    toptitle=1mm, bottomtitle=1mm,
]
\textbf{System Prompt:}

You are a strict grader. Use the rubric below to give two integer scores from 0--10:
\begin{itemize}[leftmargin=*, topsep=0pt, itemsep=0pt, parsep=2pt]
    \item \texttt{style\_score} - tweet naturalness on social media
    \item \texttt{semantic\_score} - key-information preservation
\end{itemize}

Rubric examples (do not output these examples in your answer):
\begin{verbatim}
  10 - flawless style AND all facts present
   8 - minor issues in style or small missing details
   5 - several awkward phrases OR half the facts missing
   2 - hard to read or mostly irrelevant
   0 - gibberish or unrelated to the article
\end{verbatim}
Your job is to \textbf{think step-by-step (Chain-of-Thought)} about style and semantics, then call the function \texttt{grade\_tweet} with two integer scores.

\noindent\rule{\linewidth}{0.4pt}
\textbf{Tool Schema (\texttt{grade\_tweet}):}
\begin{Verbatim}[commandchars=\\\{\}]
\textbf{Function Name}: grade_tweet
\textbf{Description}: Returns integer style and semantic scores for a tweet given a news article.

\textbf{Parameters}:
- `style_score` (integer, required): Tweet naturalness on social media, from 0 to 10.
- `semantic_score` (integer, required): Key-information preservation score, from 0 to 10.
\end{Verbatim}

\noindent\rule{\linewidth}{0.4pt}
\textbf{Example Output (Tool Call Arguments)}:
\begin{lstlisting}[style=jsonstyle]
{
  "style_score": 9,
  "semantic_score": 10
}
\end{lstlisting}

\noindent\rule{\linewidth}{0.4pt}
\textbf{Input Format:}
\begin{verbatim}
News article:
<ARTICLE>
[Original News Article Text Here]
</ARTICLE>

Generated tweet:
<TWEET>
[Style-Transferred Tweet Text Here]
</TWEET>
\end{verbatim}

\textbf{Start of Response:} The model should first provide its Chain-of-Thought reasoning, then output the tool call. The final response must include the call to the \texttt{grade\_tweet} function.

\end{tcolorbox}
\end{center}

\clearpage
\subsection{UDA Complexity Analysis}
\label{app:complexity_analysis}
To provide a comprehensive understanding of the computational overhead associated with our UDA module, we present a detailed complexity analysis of the Maximum Mean Discrepancy (\textsc{MMD}) and Kullback-Leibler (\textsc{KL}) divergence methods. This analysis examines theoretical time complexity, floating-point operations (FLOPs), and practical characteristics of GPU implementations. We assume two sets of embeddings, $E_X$ and $E_Y$, each containing $N$ samples of dimension $D$, corresponding to a batch from the source and target domains, respectively.

\subsubsection*{\textbf{Maximum Mean Discrepancy (MMD)}}
The computational core of the \textsc{MMD} loss is the construction of three kernel matrices, $K_{XX}$, $K_{YY}$, and $K_{XY}$, each of size $N \times N$.

\noindent \textbf{Time Complexity.} The calculation is dominated by the pairwise evaluation of the kernel function. For the Gaussian RBF kernel, computing the squared Euclidean distance $\|x - y\|^2$ between two $D$-dimensional vectors requires $O(D)$ operations. Since this must be done for all $N \times N$ pairs to construct each kernel matrix, the complexity of forming one matrix is $O(N^2 D)$. The final summation over the kernel matrices is $O(N^2)$. Therefore, the overall time complexity is dominated by kernel matrix computation, yielding $O(N^2 D)$.

\noindent \textbf{Floating-Point Operations (FLOPs).} We can estimate the FLOPs for the RBF kernel. The calculation of $\|x - y\|^2$ involves approximately $3D$ FLOPs ($D$ subtractions, $D$ multiplications, $D-1$ additions). Constructing the three $N \times N$ kernel matrices thus requires roughly $3 \times N^2 \times 3D = 9N^2 D$ FLOPs. This quadratic dependence on $N$ makes \textsc{MMD} computationally demanding.

\noindent \textbf{GPU Latency and Bottlenecks.} On GPU architectures, the pairwise distance calculations can be efficiently mapped to highly optimized Generalized Matrix-Matrix Multiplication (GEMM) operations. However, the $O(N^2 D)$ complexity makes the algorithm fundamentally compute-bound. Latency grows quadratically with the batch size $N$, which rapidly becomes the primary performance bottleneck. Furthermore, storing the intermediate kernel matrices requires $O(N^2)$ memory, which can exhaust the GPU's VRAM for large $N$ (e.g., $N > 4096$), thereby imposing a practical limit on the batch size.

\subsubsection*{\textbf{Kullback-Leibler (KL) Divergence}}
The \textsc{KL} divergence loss is computed element-wise on the distribution-like outputs $P$ and $Q$, which are tensors of shape $(N, D)$.

\noindent \textbf{Time Complexity.} The computation involves element-wise division, logarithm, and multiplication across two $(N, D)$ tensors, followed by a summation. Each of these element-wise operations has a time complexity proportional to the number of elements. Thus, the overall time complexity is linear with respect to both $N$ and $D$, resulting in $O(N D)$.

\noindent\textbf{Floating-Point Operations (FLOPs).} The sequence of operations (division, logarithm, multiplication, and summation) results in approximately $4ND$ FLOPs. This is substantially lower than \textsc{MMD}'s quadratic complexity.

\noindent \textbf{GPU Latency and Bottlenecks.} The element-wise nature of the \textsc{KL} divergence calculation makes it an ``embarrassingly parallel'' problem, ideally suited for GPU execution. The operations exhibit high data parallelism and low computational intensity. Consequently, the algorithm is typically memory-bound; its latency is limited not by the speed of computation but by the GPU's memory bandwidth for reading the input tensors $P$ and $Q$. The overall latency is minimal and scales linearly with the total number of elements, $N \times D$.

\subsubsection*{\textbf{Summary and Implications}}
The choice between \textsc{MMD} and \textsc{KL} divergence for domain adaptation introduces a critical trade-off in computational efficiency. Table~\ref{tab:complexity_summary} summarizes the computational characteristics of each method.

\begin{table}[h]
\centering
\caption{Computational complexity comparison of UDA methods.}
\label{tab:complexity_summary}
\rowcolors{2}{gray!10}{white} %
\begin{tabular}{l ccc}
\toprule
\textbf{Metric} & \textbf{Time Complexity} & \textbf{FLOPs (Approx.)} & \textbf{GPU Bottleneck} \\
\midrule
\textsc{MMD} & $O(N^2 D)$ & $9N^2 D$ & Compute-Bound \\
\textsc{KL} Divergence & $O(N D)$ & $4ND$ & Memory-Bound \\
\bottomrule
\end{tabular}
\end{table}

In practice, the quadratic complexity of \textsc{MMD} restricts its application to smaller batch sizes ($N$), whereas \textsc{KL} divergence remains efficient even for large-batch training. This computational difference necessitates careful management of batch sizes when using \textsc{MMD} to maintain feasible training times, while \textsc{KL} divergence offers a more scalable alternative from a computational standpoint.

\begin{table}[p]
  \centering
  \caption{Unified performance analysis of the \texttt{Qwen2.5} model family (Part 1: 0.5B, 3B, 7B). All models are evaluated from 0-shot to 200-shot settings.}
  \label{tab:unified_performance_analysis_p1}
  \resizebox{\textwidth}{!}{%
  \rowcolors{2}{gray!10}{white}
    \begin{tabular}{l rrrrrrr}
      \toprule
      & \multicolumn{3}{c}{\textbf{Classification (F1 $\uparrow$)}} & \multicolumn{2}{c}{\textbf{Political Class Detail $\uparrow$}} & \multicolumn{2}{c}{\textbf{Regression Error $\downarrow$}} \\
      \cmidrule(lr){2-4} \cmidrule(lr){5-6} \cmidrule(lr){7-8}
      \textbf{Model / Setting} & Non-political & Political & Macro Avg. & Precision & Recall & MAE & RMSE \\
      \midrule
      \multicolumn{8}{l}{\itshape\textbf{Few-Shot for \texttt{Qwen2.5-0.5B-Instruct}}} \\
      0-shot  & \pmstd{0.510}{0.022} & \pmstd{0.727}{0.007} & \pmstd{0.618}{0.014} & \pmstd{0.601}{0.007} & \pmstd{0.918}{0.006} & \pmstd{0.413}{0.007} & \pmstd{0.478}{0.005} \\
      2-shot  & \pmstd{0.568}{0.132} & \pmstd{0.717}{0.035} & \pmstd{0.643}{0.082} & \pmstd{0.629}{0.094} & \pmstd{0.855}{0.061} & \pmstd{0.297}{0.055} & \pmstd{0.364}{0.054} \\
      4-shot  & \pmstd{0.607}{0.132} & \pmstd{0.751}{0.051} & \pmstd{0.679}{0.089} & \pmstd{0.662}{0.126} & \pmstd{0.901}{0.088} & \pmstd{0.330}{0.031} & \pmstd{0.395}{0.038} \\
      6-shot  & \pmstd{0.577}{0.098} & \pmstd{0.750}{0.032} & \pmstd{0.664}{0.062} & \pmstd{0.622}{0.040} & \pmstd{0.951}{0.052} & \pmstd{0.305}{0.016} & \pmstd{0.367}{0.018} \\
      8-shot  & \pmstd{0.525}{0.233} & \pmstd{0.753}{0.058} & \pmstd{0.639}{0.143} & \pmstd{0.628}{0.088} & \pmstd{0.960}{0.067} & \pmstd{0.257}{0.024} & \pmstd{0.317}{0.032} \\
      10-shot & \pmstd{0.619}{0.099} & \pmstd{0.773}{0.023} & \pmstd{0.696}{0.061} & \pmstd{0.647}{0.053} & \pmstd{0.970}{0.046} & \pmstd{0.287}{0.056} & \pmstd{0.349}{0.062} \\
      20-shot & \pmstd{0.824}{0.090} & \pmstd{0.867}{0.046} & \pmstd{0.845}{0.068} & \pmstd{0.797}{0.097} & \pmstd{0.964}{0.038} & \pmstd{0.285}{0.019} & \pmstd{0.357}{0.015} \\
      50-shot & \pmstd{0.896}{0.063} & \pmstd{0.916}{0.043} & \pmstd{0.906}{0.053} & \pmstd{0.849}{0.074} & \pmstd{0.999}{0.001} & \pmstd{0.282}{0.027} & \pmstd{0.352}{0.027} \\
      100-shot& \pmstd{0.988}{0.007} & \pmstd{0.988}{0.006} & \pmstd{0.988}{0.006} & \pmstd{0.980}{0.016} & \pmstd{0.996}{0.006} & \pmstd{0.274}{0.018} & \pmstd{0.338}{0.017} \\
      200-shot& \pmstd{0.988}{0.009} & \pmstd{0.988}{0.008} & \pmstd{0.988}{0.009} & \pmstd{0.978}{0.018} & \pmstd{0.998}{0.003} & \pmstd{0.245}{0.023} & \pmstd{0.309}{0.027} \\
      \midrule
      \multicolumn{8}{l}{\itshape\textbf{Few-Shot for \texttt{Qwen2.5-3B-Instruct}}} \\
      0-shot  & \pmstd{0.946}{0.003} & \pmstd{0.943}{0.003} & \pmstd{0.945}{0.003} & \pmstd{0.945}{0.008} & \pmstd{0.942}{0.006} & \pmstd{0.283}{0.005} & \pmstd{0.333}{0.005} \\
      2-shot  & \pmstd{0.977}{0.004} & \pmstd{0.976}{0.005} & \pmstd{0.977}{0.005} & \pmstd{0.987}{0.005} & \pmstd{0.965}{0.012} & \pmstd{0.258}{0.016} & \pmstd{0.299}{0.021} \\
      4-shot  & \pmstd{0.970}{0.015} & \pmstd{0.968}{0.018} & \pmstd{0.969}{0.016} & \pmstd{0.984}{0.025} & \pmstd{0.954}{0.042} & \pmstd{0.271}{0.035} & \pmstd{0.321}{0.046} \\
      6-shot  & \pmstd{0.981}{0.012} & \pmstd{0.979}{0.014} & \pmstd{0.980}{0.013} & \pmstd{0.994}{0.006} & \pmstd{0.966}{0.027} & \pmstd{0.260}{0.021} & \pmstd{0.312}{0.029} \\
      8-shot  & \pmstd{0.984}{0.008} & \pmstd{0.982}{0.010} & \pmstd{0.983}{0.009} & \pmstd{0.993}{0.004} & \pmstd{0.972}{0.022} & \pmstd{0.275}{0.024} & \pmstd{0.335}{0.032} \\
      10-shot & \pmstd{0.980}{0.014} & \pmstd{0.978}{0.016} & \pmstd{0.979}{0.015} & \pmstd{0.997}{0.003} & \pmstd{0.961}{0.032} & \pmstd{0.268}{0.030} & \pmstd{0.331}{0.038} \\
      20-shot & \pmstd{0.987}{0.002} & \pmstd{0.986}{0.002} & \pmstd{0.986}{0.002} & \pmstd{0.998}{0.003} & \pmstd{0.974}{0.004} & \pmstd{0.260}{0.028} & \pmstd{0.325}{0.036} \\
      50-shot & \pmstd{0.985}{0.007} & \pmstd{0.983}{0.009} & \pmstd{0.984}{0.008} & \pmstd{1.000}{0.000} & \pmstd{0.967}{0.018} & \pmstd{0.291}{0.034} & \pmstd{0.368}{0.032} \\
      100-shot& \pmstd{0.978}{0.005} & \pmstd{0.972}{0.009} & \pmstd{0.975}{0.006} & \pmstd{1.000}{0.000} & \pmstd{0.947}{0.017} & \pmstd{0.283}{0.025} & \pmstd{0.367}{0.029} \\
      200-shot& \pmstd{0.992}{0.002} & \pmstd{0.989}{0.003} & \pmstd{0.990}{0.002} & \pmstd{0.998}{0.002} & \pmstd{0.980}{0.006} & \pmstd{0.273}{0.015} & \pmstd{0.353}{0.014} \\
      \midrule
      \multicolumn{8}{l}{\itshape\textbf{Few-Shot for \texttt{Qwen2.5-7B-Instruct}}} \\
      0-shot  & \pmstd{0.927}{0.002} & \pmstd{0.905}{0.003} & \pmstd{0.916}{0.002} & \pmstd{0.992}{0.003} & \pmstd{0.833}{0.004} & \pmstd{0.238}{0.004} & \pmstd{0.288}{0.005} \\
      2-shot  & \pmstd{0.936}{0.021} & \pmstd{0.928}{0.026} & \pmstd{0.932}{0.024} & \pmstd{0.986}{0.005} & \pmstd{0.879}{0.049} & \pmstd{0.242}{0.005} & \pmstd{0.292}{0.007} \\
      4-shot  & \pmstd{0.920}{0.032} & \pmstd{0.905}{0.044} & \pmstd{0.912}{0.038} & \pmstd{0.986}{0.006} & \pmstd{0.840}{0.079} & \pmstd{0.230}{0.007} & \pmstd{0.282}{0.012} \\
      6-shot  & \pmstd{0.918}{0.024} & \pmstd{0.903}{0.035} & \pmstd{0.911}{0.030} & \pmstd{0.987}{0.008} & \pmstd{0.835}{0.063} & \pmstd{0.243}{0.007} & \pmstd{0.301}{0.007} \\
      8-shot  & \pmstd{0.906}{0.024} & \pmstd{0.886}{0.035} & \pmstd{0.896}{0.029} & \pmstd{0.990}{0.007} & \pmstd{0.804}{0.061} & \pmstd{0.246}{0.007} & \pmstd{0.305}{0.011} \\
      10-shot & \pmstd{0.898}{0.019} & \pmstd{0.874}{0.028} & \pmstd{0.886}{0.024} & \pmstd{0.995}{0.002} & \pmstd{0.781}{0.047} & \pmstd{0.236}{0.012} & \pmstd{0.291}{0.015} \\
      20-shot & \pmstd{0.923}{0.031} & \pmstd{0.908}{0.043} & \pmstd{0.916}{0.037} & \pmstd{0.995}{0.002} & \pmstd{0.838}{0.074} & \pmstd{0.244}{0.016} & \pmstd{0.306}{0.019} \\
      50-shot & \pmstd{0.961}{0.014} & \pmstd{0.958}{0.015} & \pmstd{0.960}{0.014} & \pmstd{0.996}{0.003} & \pmstd{0.924}{0.031} & \pmstd{0.241}{0.018} & \pmstd{0.304}{0.020} \\
      100-shot& \pmstd{0.964}{0.008} & \pmstd{0.962}{0.009} & \pmstd{0.963}{0.009} & \pmstd{0.998}{0.000} & \pmstd{0.929}{0.017} & \pmstd{0.237}{0.010} & \pmstd{0.307}{0.012} \\
      200-shot& \pmstd{0.974}{0.008} & \pmstd{0.973}{0.008} & \pmstd{0.973}{0.008} & \pmstd{0.998}{0.002} & \pmstd{0.950}{0.017} & \pmstd{0.243}{0.012} & \pmstd{0.315}{0.018} \\
      \bottomrule
    \end{tabular}%
  }%
\end{table}

\begin{table}[p]
  \centering
  \caption{Unified performance analysis of the \texttt{Qwen2.5} model family (Part 2: 14B, 32B, 72B). All models are evaluated from 0-shot to 200-shot settings.}
  \label{tab:unified_performance_analysis_p2}
  \resizebox{\linewidth}{!}{%
  \rowcolors{2}{gray!10}{white}
    \begin{tabular}{l rrrrrrr}
      \toprule
      & \multicolumn{3}{c}{\textbf{Classification (F1 $\uparrow$)}} & \multicolumn{2}{c}{\textbf{Political Class Detail $\uparrow$}} & \multicolumn{2}{c}{\textbf{Regression Error $\downarrow$}} \\
      \cmidrule(lr){2-4} \cmidrule(lr){5-6} \cmidrule(lr){7-8}
      \textbf{Model / Setting} & Non-political & Political & Macro Avg. & Precision & Recall & MAE & RMSE \\
      \midrule
      \multicolumn{8}{l}{\itshape\textbf{Few-Shot for \texttt{Qwen2.5-14B-Instruct}}} \\
      0-shot  & \pmstd{0.918}{0.003} & \pmstd{0.900}{0.004} & \pmstd{0.909}{0.003} & \pmstd{0.986}{0.001} & \pmstd{0.827}{0.006} & \pmstd{0.227}{0.004} & \pmstd{0.285}{0.003} \\
      2-shot  & \pmstd{0.949}{0.022} & \pmstd{0.941}{0.029} & \pmstd{0.945}{0.026} & \pmstd{0.992}{0.003} & \pmstd{0.896}{0.054} & \pmstd{0.219}{0.007} & \pmstd{0.273}{0.011} \\
      4-shot  & \pmstd{0.955}{0.009} & \pmstd{0.949}{0.011} & \pmstd{0.952}{0.010} & \pmstd{0.994}{0.001} & \pmstd{0.908}{0.021} & \pmstd{0.216}{0.002} & \pmstd{0.271}{0.007} \\
      6-shot  & \pmstd{0.963}{0.014} & \pmstd{0.959}{0.017} & \pmstd{0.961}{0.016} & \pmstd{0.991}{0.004} & \pmstd{0.929}{0.034} & \pmstd{0.235}{0.025} & \pmstd{0.299}{0.032} \\
      8-shot  & \pmstd{0.961}{0.019} & \pmstd{0.956}{0.024} & \pmstd{0.959}{0.022} & \pmstd{0.992}{0.005} & \pmstd{0.924}{0.047} & \pmstd{0.232}{0.008} & \pmstd{0.295}{0.014} \\
      10-shot & \pmstd{0.962}{0.015} & \pmstd{0.958}{0.018} & \pmstd{0.960}{0.017} & \pmstd{0.988}{0.003} & \pmstd{0.930}{0.036} & \pmstd{0.236}{0.009} & \pmstd{0.305}{0.013} \\
      20-shot & \pmstd{0.980}{0.006} & \pmstd{0.978}{0.006} & \pmstd{0.979}{0.006} & \pmstd{0.990}{0.002} & \pmstd{0.967}{0.013} & \pmstd{0.235}{0.010} & \pmstd{0.300}{0.015} \\
      50-shot & \pmstd{0.982}{0.002} & \pmstd{0.981}{0.002} & \pmstd{0.981}{0.002} & \pmstd{0.991}{0.002} & \pmstd{0.971}{0.005} & \pmstd{0.239}{0.006} & \pmstd{0.309}{0.011} \\
      100-shot& \pmstd{0.968}{0.007} & \pmstd{0.965}{0.008} & \pmstd{0.966}{0.008} & \pmstd{0.988}{0.004} & \pmstd{0.943}{0.015} & \pmstd{0.241}{0.007} & \pmstd{0.311}{0.006} \\
      200-shot& \pmstd{0.979}{0.004} & \pmstd{0.979}{0.005} & \pmstd{0.979}{0.004} & \pmstd{0.995}{0.001} & \pmstd{0.963}{0.009} & \pmstd{0.223}{0.004} & \pmstd{0.301}{0.004} \\
      \midrule
      \multicolumn{8}{l}{\itshape\textbf{Few-Shot for \texttt{Qwen2.5-32B-Instruct}}} \\
      0-shot  & \pmstd{0.946}{0.001} & \pmstd{0.937}{0.002} & \pmstd{0.942}{0.002} & \pmstd{0.991}{0.001} & \pmstd{0.889}{0.004} & \pmstd{0.225}{0.002} & \pmstd{0.283}{0.001} \\
      2-shot  & \pmstd{0.956}{0.008} & \pmstd{0.955}{0.009} & \pmstd{0.956}{0.008} & \pmstd{0.993}{0.002} & \pmstd{0.921}{0.017} & \pmstd{0.206}{0.009} & \pmstd{0.263}{0.012} \\
      4-shot  & \pmstd{0.962}{0.008} & \pmstd{0.962}{0.008} & \pmstd{0.962}{0.008} & \pmstd{0.994}{0.001} & \pmstd{0.933}{0.015} & \pmstd{0.214}{0.005} & \pmstd{0.275}{0.006} \\
      6-shot  & \pmstd{0.968}{0.005} & \pmstd{0.969}{0.005} & \pmstd{0.968}{0.005} & \pmstd{0.992}{0.003} & \pmstd{0.946}{0.011} & \pmstd{0.212}{0.005} & \pmstd{0.272}{0.008} \\
      8-shot  & \pmstd{0.976}{0.006} & \pmstd{0.977}{0.006} & \pmstd{0.976}{0.006} & \pmstd{0.995}{0.002} & \pmstd{0.959}{0.010} & \pmstd{0.210}{0.007} & \pmstd{0.272}{0.011} \\
      10-shot & \pmstd{0.973}{0.009} & \pmstd{0.974}{0.010} & \pmstd{0.973}{0.009} & \pmstd{0.994}{0.001} & \pmstd{0.954}{0.018} & \pmstd{0.207}{0.006} & \pmstd{0.270}{0.010} \\
      20-shot & \pmstd{0.980}{0.004} & \pmstd{0.980}{0.004} & \pmstd{0.980}{0.004} & \pmstd{0.993}{0.002} & \pmstd{0.968}{0.009} & \pmstd{0.219}{0.011} & \pmstd{0.284}{0.016} \\
      50-shot & \pmstd{0.989}{0.002} & \pmstd{0.989}{0.002} & \pmstd{0.989}{0.002} & \pmstd{0.995}{0.002} & \pmstd{0.984}{0.004} & \pmstd{0.218}{0.007} & \pmstd{0.285}{0.008} \\
      100-shot& \pmstd{0.990}{0.001} & \pmstd{0.990}{0.001} & \pmstd{0.990}{0.001} & \pmstd{0.996}{0.001} & \pmstd{0.985}{0.002} & \pmstd{0.225}{0.007} & \pmstd{0.296}{0.007} \\
      200-shot& \pmstd{0.992}{0.002} & \pmstd{0.993}{0.002} & \pmstd{0.992}{0.002} & \pmstd{0.996}{0.001} & \pmstd{0.990}{0.003} & \pmstd{0.212}{0.011} & \pmstd{0.284}{0.011} \\
      \midrule
      \multicolumn{8}{l}{\itshape\textbf{Few-Shot for \texttt{Qwen2.5-72B-Instruct}}} \\
      0-shot  & \pmstd{0.955}{0.002} & \pmstd{0.950}{0.003} & \pmstd{0.953}{0.002} & \pmstd{0.994}{0.002} & \pmstd{0.910}{0.004} & \pmstd{0.216}{0.004} & \pmstd{0.275}{0.004} \\
      2-shot  & \pmstd{0.946}{0.015} & \pmstd{0.935}{0.020} & \pmstd{0.940}{0.017} & \pmstd{0.990}{0.003} & \pmstd{0.887}{0.037} & \pmstd{0.206}{0.004} & \pmstd{0.265}{0.003} \\
      4-shot  & \pmstd{0.972}{0.007} & \pmstd{0.969}{0.009} & \pmstd{0.971}{0.008} & \pmstd{0.989}{0.003} & \pmstd{0.950}{0.018} & \pmstd{0.206}{0.009} & \pmstd{0.268}{0.011} \\
      6-shot  & \pmstd{0.967}{0.012} & \pmstd{0.962}{0.014} & \pmstd{0.964}{0.013} & \pmstd{0.989}{0.002} & \pmstd{0.937}{0.029} & \pmstd{0.212}{0.008} & \pmstd{0.277}{0.011} \\
      8-shot  & \pmstd{0.972}{0.010} & \pmstd{0.969}{0.012} & \pmstd{0.971}{0.011} & \pmstd{0.990}{0.003} & \pmstd{0.949}{0.022} & \pmstd{0.225}{0.016} & \pmstd{0.292}{0.022} \\
      10-shot & \pmstd{0.965}{0.014} & \pmstd{0.959}{0.017} & \pmstd{0.962}{0.015} & \pmstd{0.993}{0.002} & \pmstd{0.928}{0.032} & \pmstd{0.217}{0.018} & \pmstd{0.285}{0.022} \\
      20-shot & \pmstd{0.974}{0.006} & \pmstd{0.971}{0.007} & \pmstd{0.972}{0.006} & \pmstd{0.992}{0.002} & \pmstd{0.950}{0.015} & \pmstd{0.214}{0.011} & \pmstd{0.279}{0.013} \\
      50-shot & \pmstd{0.988}{0.005} & \pmstd{0.988}{0.006} & \pmstd{0.988}{0.006} & \pmstd{0.991}{0.002} & \pmstd{0.985}{0.012} & \pmstd{0.213}{0.006} & \pmstd{0.284}{0.009} \\
      100-shot& \pmstd{0.987}{0.003} & \pmstd{0.992}{0.001} & \pmstd{0.989}{0.002} & \pmstd{0.992}{0.001} & \pmstd{0.993}{0.003} & \pmstd{0.220}{0.014} & \pmstd{0.297}{0.016} \\
      200-shot& \pmstd{0.985}{0.003} & \pmstd{0.993}{0.001} & \pmstd{0.989}{0.002} & \pmstd{0.994}{0.002} & \pmstd{0.991}{0.002} & \pmstd{0.212}{0.007} & \pmstd{0.290}{0.008} \\
      \bottomrule
    \end{tabular}%
  }%
\end{table}